\documentclass[11pt,preprint]{aastex} 
  
\shorttitle{} 
\shortauthors{} 
 
\begin{document} 
  
\title{Thermal and Fragmentation Properties of Star-forming Clouds \\ 
in Low-metallicity Environments} 
\author{K. Omukai \altaffilmark{1},  
T. Tsuribe \altaffilmark{2},  
R. Schneider \altaffilmark{3} and  
A. Ferrara \altaffilmark{4}} 
\altaffiltext{1}{National Astronomical Observatory of Japan,  
Mitaka, Tokyo 181-8588, Japan; omukai@th.nao.ac.jp} 
\altaffiltext{2}{Department of Earth and Space Science,  
Osaka University, Toyonaka 606-5504, Japan;  
tsuribe@vega.ess.sci.osaka-u.ac.jp} 
\altaffiltext{3}{INAF--Osservatorio Astrofisico di Arcetri,  
Largo E. Fermi 5, 50125 Florence, Italy} 
\altaffiltext{4}{SISSA/International School for Advanced Studies,  
Via Beirut 4, 34100 Trieste, Italy} 
 
\begin{abstract} 
The thermal and chemical evolution of star-forming clouds is  
studied for different gas metallicities, $Z$, using the model of Omukai (2000), updated to include deuterium chemistry and the effects of cosmic microwave background (CMB) radiation. HD-line cooling dominates the thermal balance of clouds when  $Z \sim 10^{-5}-10^{-3} Z_{\sun}$ and density $\approx 10^{5} {\rm cm^{-3}}$. Early on, CMB radiation prevents the gas temperature to fall below $T_{CMB}$, although this hardly alters the cloud thermal evolution in low-metallicity gas.  
From the derived temperature evolution, we assess cloud/core fragmentation as a function of metallicity from linear perturbation theory, which requires that the core elongation ${\cal E} \equiv (b-a)/a > {\cal E}_{\rm NL} \sim 1$, where $a$ ($b$) is the short (long) core axis length. The fragment mass is given  
by the thermal Jeans mass at ${\cal E} ={\cal E}_{\rm NL}$. 
Given these assumptions and the initial (gaussian) distribution of ${\cal E}$ 
we compute the fragment mass distribution as a function of metallicity.  
We find that: (i) For $Z=0$, all fragments are very massive, $\la 10^{3}M_{\sun}$, consistently with previous studies; (ii) for $Z>10^{-6} Z_\sun$ a few clumps go through an additional high density ($\ga 10^{10}{\rm cm^{-3}}$) fragmentation phase driven by dust-cooling, leading to low-mass fragments; 
(iii) The mass fraction in low-mass fragments is initially very small, but at $Z \sim 10^{-5}Z_{\sun}$ it becomes dominant and continues to grow as $Z$ is increased; (iv) as a result of the two fragmentation modes, a bimodal   
mass distribution emerges in $0.01 < Z/Z_{\sun} < 0.1$. 
(v) For $\ga 0.1Z_{\sun}$, the two peaks merge into a singly-peaked mass function which might be regarded as the precursor of the ordinary Salpeter-like IMF.  
\end{abstract}  
 
\keywords{cosmology: theory --- galaxies: formation --- stars: formation} 
  
\section{Introduction} 
The presence of a massive stellar population in the early universe  
has been frequently invoked to account for some recent observations.  
Most notably the large electron scattering optical depth of  
$\tau_{\rm e} = 0.17^{+0.08}_{-0.07} $  
detected by the WMAP satellite, corresponding to the reionization of 
the intergalactic medium at redshifts $\la 10$  
(Spergel et al. 2003; Kogut et al. 2003),  
is interpreted with photoionization by massive population III (Pop III) stars. 
In addition, the amplitude and spectral break found at 1 $\mu$m in the infrared  
background radiation is explained by radiation from Pop III stars at $z>9$   
(Santos, Bromm, \& Kamionkowski 2002; Salvaterra \& Ferrara 2003). 
 
Even before those evidences, theorists claimed that the  
first stars in the universe should be very massive, typically  
$100-1000M_{\sun}$ 
(e.g., Abel, Bryan, \& Norman 2000, 2002; Omukai \& Palla 2001, 2003).  
Among them, Bromm et al. (1999, 2002) investigated the collapse  
and fragmentation of primordial-gas clouds with 3-dimensional  
numerical hydrodynamics. 
They discovered that fragmentation occurs at density  
$\sim 10^{4} {\rm cm^{-3}}$.  
As a result of this, dense cores with mass-scale  
$\sim 1000M_{\sun}$ are produced, which do not fragment  
into smaller sub-clumps after that. 
This fragmentation history can be interpreted from the view point of  
thermal evolution of primordial gas. 
In low-temperature ($\la 10^{4}$K) primordial gas, the only important  
cooling agent is rotational/vibrational line emission by molecular hydrogen. 
Once a sufficient amount of H$_2$ for dynamical collapse is formed,  
resultant cooling causes a temperature decrease with increasing density. 
The collapse in this phase is so rapid that density perturbations have not  
enough time to grow and the clouds do not fragment. 
However, at $\sim 10^{4} {\rm cm^{-3}}$ the rotational levels of  
H$_2$ reach local thermodynamic equilibrium, and the cooling becomes 
inefficient at higher densities.  
Now the temperature begins to increase with increasing density. 
The transition from temperature decrease to increase causes the clouds to   
remain for long time intervals (or to ``loiter'') at $\sim 10^{4} {\rm cm^{-3}}$  
at 200K, the minimum temperature needed for excitation of H$_2$ levels (the first excited level locates at $\simeq 600$K).  
At this moment, as there is sufficient time for fluctuations to grow, fragmentation occurs. 
The fragmentation mass-scale is set by the thermal Jeans mass at this time,  
i.e., $\sim 1000M_{\sun}$, using the number density   
$\sim 10^{4} {\rm cm^{-3}}$ and temperature $\sim 200$K. 
Those massive dense cores are expected to grow eventually to massive stars 
because even though protostars are initially small, rapid accretion onto  
the formed protostars is not halted until at least it grows to  $\sim 100M_{\sun}$ 
without dust grains (Omukai \& Palla 2003; Tan \& McKee 2004; 
Bromm \& Loeb 2004). 
 
Along the same lines, the characteristic mass-scale ($0.1-1M_{\sun}$) of present-day Galactic star formation may be derived (Larson 2004). 
In the high-mass regime ($\la M_{\sun}$), the number of field stars decreases  
with increasing stellar mass following the Salpeter IMF, 
while the slope of the IMF flattens in the sub-solar mass regime  
and eventually decreases for stars below about $0.1M_{\sun}$. 
The precise origin of this mass-scale is still obscure, but Larson (2004)  
interpreted it as corresponding to the Jeans mass at the  
end of rapid cooling phase caused by dust thermal emission. 
 
From the arguments above, a close relation between the thermal evolution  
and fragmentation is evident. 
Fragmentation occurs at the end of temperature-decreasing phase,  
or in other words, when the isothermality breaks, namely the effective  
ratio of specific heat $\gamma \equiv {\rm dlog}p/{\rm dlog} \rho > 1$; 
the typical fragmentation mass-scale is given by the thermal Jeans mass  
at that epoch. 
 
Significance of the role played by the massive stellar population  
in early universe crucially depends on the duration of massive-star  
formation epoch and the amount of massive stars formed. 
As metal enrichment proceeds in interstellar medium (ISM),  
it might affect the stellar mass-scale in two different ways. 
First, pre-stellar thermal evolution is altered,  
and consequently the typical mass-scale of fragmentation changes  
with metallicity.  
Second, in sufficiently high dust-to-gas ratio cases,  
i.e. high-metallicity environment,  
the accretion onto massive protostars is inhibited by radiation force  
onto dust grains resulting in the scarcity of massive stars.  
As for the former effect, even with metallicities as small as  
$10^{-5\pm 1} Z_{\sun}$ the thermal evolution is altered 
significantly, and low-mass fragments are likely to be produced  
(Schneider et al. 2002; 2003). 
As for the latter, a metallicity as high as $\sim 10^{-2}Z_{\sun}$ is  
necessary to affect the accretion process by radiation force 
onto dust grains, if the dust opacity for a given metallicity remains  
the same as in local ISM (Omukai \& Palla 2003). 
Since the former effect becomes important earlier in the  
history of metal-enrichment, the transition from massive to low-mass star  
formation mode is probably caused by the fragmentation process rather than  
by halting the accretion. 
 
Based on the thermal evolution obtained by Omukai (2000; Paper I), 
Schneider et al. (2002, 2003) discussed fragmentation properties  
for low-metallicity clouds and pointed out that even within the  metallicity  
range $10^{-5\pm 1} Z_{\sun}$ low-mass fragmentation is possible.  
However, the thermal evolution of those low-metallicity clouds  
is rather complicated.  
The molecular cooling phase is followed by temperature  
increase owing to H$_2$ formation, and finally by  
dust-induced temperature drop before becoming optically thick. 
That is, there are a few up-and-downs in temperature evolution 
and the epoch of isothermality break is not unique. 
Therefore, it is difficult to determine whether fragmentation indeed   
occurs. 
In particular, after the first fragmentation in molecular-cooling phase,  
the cloud starts run-away collapse, during which density 
fluctuations might be erased by pressure force. 
If so, with little elongation of the core, the fragmentation at higher density  
would be prohibited. Therefore, whether fragmentation occurs in the 
dust-cooling phase  is still unclear.  
 
To clarify this point, here we follow the evolution of elongation of  
dense cores during the run-away collapse and discuss when the final  
fragmentation occurs. 
Our results confirm that  fragmentation does occur  
in the dust-cooling phase and sub-solar mass fragments are indeed produced. 
In some cases, dust-induced fragmentation occurs even in metallicity  
as low as $10^{-6}Z_{\sun}$. 
We have improved the previous model (Paper I) by including deuterium  
chemistry. Although HD cooling is important neither in the metal-free case nor in  
metal-rich cases, HD is found to be the dominant coolant  
in cases with slight ($\sim 10^{-4}Z_{\sun}$) metallicity  
for some density ranges. 
Finally, we have constructed reduced chemical models that
can reproduce the thermal evolution in an approximate way and
might be used to include the chemical network of metals
in future challenging simulations. 
 
The paper is organized as follows. 
In \S 2, our model for star-forming clouds is briefly described with  
emphasis on modification from the previous one. 
In \S 3, the calculated thermal and chemical evolution is presented. 
Finally in \S 4, we summarize and discuss the results with some  
speculations on the evolution of IMF. 
In Appendix, by applying a simple model for fragmentation to 
the thermal evolution derived in \S 3, 
the fragmentation mass-scales are discussed.
 
\section{Model for Pre-stellar Clouds} 
Our analysis in this paper is based on the model developed in paper I.  
Prestellar clouds are treated by a one-zone model where involved physical  
variables are regarded as those at the center. 
Thermal and chemical variables at the center,  
e.g., temperature $T=T(n_{\rm H})$, concentration of chemical species $i$  
$y(i)=y(i)(n_{\rm H})$\footnote{The concentration $y(i)$ of the chemical  
species $i$ is defined by $y(i)=n(i)/n_{\rm H}$, where n($i$) is  
the number density of the species $i$.} etc., are followed as a  
function of the number density of hydrogen nuclei $n_{\rm H}$. 
While detailed radiative and chemical processes are included in this model,  
the dynamics of the clouds is largely simplified. 
 
The temperature evolution is followed by solving the energy equation: 
\begin{equation} 
\frac{de}{dt}=-p \frac{d}{dt} \frac{1}{\rho}- {\Lambda}_{\rm net},   
\label{eq:energy}  
\end{equation} 
where the pressure 
\begin{equation} 
p=\frac{\rho k T}{\mu m_{\rm H}} 
\end{equation} 
and the specific thermal energy is written as  
\begin{equation} 
e=\frac{1}{\gamma_{\rm ad} -1}\frac{k T}{ \mu m_{\rm H}} 
\label{eq:defen} 
\end{equation} 
by using the adiabatic exponent $\gamma_{\rm ad}$, the temperature $T$,  
and mean molecular weight $\mu$. 
In equation (\ref{eq:energy}), the first term on the right hand side  
is the compressional heating,  
and the second term ${\Lambda}_{\rm net}$ is the net cooling due to  
the radiative processes and chemical reactions.  
The net cooling rate consists of three components: 
\begin{equation} 
{\Lambda}_{\rm net}= \Lambda_{\rm line} + \Lambda_{\rm cont} 
+ \Lambda_{\rm chem},  
\end{equation} 
i.e., the cooling rate owing to line radiation $\Lambda_{\rm line}$,  
to continuum radiation $\Lambda_{\rm cont}$, and to chemical reactions 
$\Lambda_{\rm chem}$. 
The last term can be negative for chemical heating rather than cooling. 
In our treatment, as in equation (\ref{eq:defen}),  
the specific thermal energy does not include chemical binding energy. 
Variations in chemical binding energy are regarded as cooling and  
heating processes and included in the net cooling rate  
${\Lambda}_{\rm net}$. 
As the line radiation process, we include molecular line emission of  
H$_2$, HD, OH and H$_2$O, and atomic fine-structure line emission of  
[CI], [CII], [OI]. 
The H$_2$ collisional transition rates of low-lying rotational levels ($J<5$)  
are updated by following Galli \& Palla (1998) from paper I where  
Hollenbach \& McKee (1979)'s coefficients were used. 
This modification lowers the H$_2$ cooling rate below 300K  
and alters the behavior around $n_{\rm H} \sim 10^{4} {\rm cm^{-3}}$  
for low-metallicity clouds. 
The HD parameters are from Galli \& Palla (1998)'s compilation. 
Other parameters are the same as in paper I. 
Continuum cooling consists of that by dust thermal emission  
and by H$_2$ collision-induced emission. 
In optically thick cases, cooling rate is reduced by multiplying  
by the escape probability of photons.  
Unless otherwise stated, the treatment of these processes are the same  
as in paper I. 
 
External radiation is not included in paper I.  
Cosmic microwave background radiation (CMB) in early universe,  
whose temperature is as high as several tens K,  
can significantly influence the thermal evolution of star-forming clouds.  
Although there should exist additional radiation field from stars or quasars, 
the amount and spectrum are quite uncertain. 
Therefore, in this paper, we consider only CMB as an external radiation  
source. 
CMB enters in our model in two ways; first through the level population  
of atomic/molecular lines, and second through the heating of dust grains. 
These effects prevent the gas and dust temperature to fall below  
the radiation temperature. 
The former effect is included in calculating level  
population of atomic and molecular species as in Omukai(2001). 
The latter effect is included in determining the grain temperature by  
the energy balance equation of grains: 
\begin{equation} 
4 \sigma T_{\rm gr}^{4} \kappa_{\rm gr}= \Lambda_{\rm gas \rightarrow dust} 
+4 \sigma T_{\rm rad}^{4} \kappa_{\rm gr}, 
\label{eq:grtemp} 
\end{equation} 
where $\Lambda_{\rm gas-dust}$ is energy transfer rate from gas to dust  
per unit mass, which is given by equation (19) of paper I  
\footnote{This equation contains a typographical error. The correct  
numerical factor is $1.1 \times 10^{-5}$ instead of $1.1 \times 10^{5}$}. 
The left hand side of equation (\ref{eq:grtemp}) is the cooling rate of  
grains and the right hand side is heating rate by collisions with gas  
particles (the first term) and by CMB heating (the second term). 
Here we use the approximation that the Planck mean opacity for absorption  
is the same as that for emission. 
This is a good approximation, being exact when 
$T_{\rm gr}=T_{\rm rad}$, i.e. when CMB effect is significant. 
 
In paper I, we have solved H, He, C, and O chemical network  
among 50 species and 478 reactions. 
In this work, we consider the additional contributions of deuterium  
chemistry and HD cooling. 
The deuterium chemical network and the reaction rates are  
taken from Table 1 of Nakamura \& Umemura (2002), which include 
18 reactions among 5 species (${\rm D}$, ${\rm D^{+}}$, ${\rm D^{-}}$,  
${\rm HD}$, ${\rm HD^{+}}$). 
Deuterium abundance is $y_{\rm D}=3 \times 10^{-5}$. 
For gas metallicity  equal to that of the solar neighborhood  
$Z_{\rm local}$,  
gas-phase elemental abundances of carbon and oxygen are  
$y_{\rm C, gas}=0.927 \times 10^{-4}$ and  
$y_{\rm O, gas}=3.568 \times 10^{-4}$. 
The mass fraction of grains is $0.939 \times 10^{-2}$ below the water-ice  
evaporation temperature, and it decreases for higher temperatures  
as, following the grain model by Pollack et al. (1994), 
each grain component evaporates. 
For lower gas metallicities, these values are reduced proportionally. 
We denote the relative abundances relative to the solar  
neighborhood values by $[Z/H] \equiv {\rm log}_{10}(Z/Z_{\rm local})$. 
 
Next, we describe our prescription for the dynamics. 
In our model, the only two quantities related to the dynamics are  
the collapse timescale and the optical depth. 
Since we are interested in the evolution during the run-away collapse 
phase, we assume that the dynamics is described by the  
Larson-Penston-type self-similar solution, which reproduces the  
dynamics of the run-away collapse phase very well 
(Larson 1969; Penston 1969).\footnote{
The solution was found by Larson (1969) and Penston (1969) 
for the isothermal case and later extended to the polytropic cases 
by Yahil(1983). }
Given the collapse timescale $t_{\rm col}$ of the self-similar solution,  
the evolution of the central density is followed by 
\begin{equation} 
\frac{d \rho}{dt}=\frac{\rho}{t_{\rm col}}.  
\end{equation} 
The collapse timescale of the self-similar solutions is given by  
\begin{equation} 
t_{\rm col}=\frac{1}{\sqrt{1-f}} t_{\rm col, 0} 
\end{equation} 
where $f$ is the ratio between the pressure gradient force and gravity  
at the center and  
\begin{equation} 
t_{\rm col, 0}=\sqrt{\frac{3 \pi}{32 G \rho}} 
\end{equation} 
is the collapse timescale for  pressure-free collapse. 
We approximate the ratio of pressure to gravity $f$ 
in the central region of the self-similar solutions  
by the following formula:  
\begin{equation} 
f= \left\{  
\begin{array}{lc} 
 0 
& \gamma < 0.83\\ 
0.6+2.5(\gamma-1)-6.0(\gamma-1)^2 
& 0.83< \gamma < 1\\ 
1.0+0.2(\gamma-4/3)-2.9(\gamma-4/3)^2 
& \gamma > 1.  
\end{array} 
 \right. 
\label{eq:abn} 
\end{equation}
Error in $f$ is less than 0.05. 
At $\gamma=4/3$, $f$ reaches unity and $t_{\rm col}$ diverges. 
This indicates a break-down of our assumption that  
hydrodynamical evolution is approximated by the self-similar solution.  
In reality, a hydrostatic core is formed at this point. 
Thereafter the core contracts as its mass increases owing to accretion  
of the envelope. 
To mimic this evolution, we set an arbitrary upper limit of  $f = 0.95$. 
The value of this upper limit does not affects the thermal evolution  
significantly unless when it is very close to unity, say $f \ga 0.99$. 
Early in the evolution, before the clouds experience the first  
fragmentation, the dynamics has not yet converged to the run-away collapse. 
In this regime, we assumed free-fall collapse, i.e., $f=0$. 
 
The density distribution of the self-similar solution,  
which is needed to estimate the optical depth, is as follows: 
inside the central region, whose length scale is roughly given by thermal  
Jeans length 
\begin{equation}
\lambda_{\rm J}=c_{\rm s} \sqrt{\pi / G \rho},
\end{equation}
where $c_{\rm s}$ is the sound speed, 
the density is approximately spatially constant, 
while in the envelope ($r \ga \lambda_{\rm J}$), it decreases  
outward as $\rho \propto r^{-2/(2-\gamma)}$. 
Since the column density in the envelope is at most the same as that  
of the central region and, in addition,  the large Doppler shift  
owing to bulk velocity reduces the optical depth of the lines, we consider only  
the shielding of photons by material in the central region. 
Namely, if the cloud becomes optically thick, the optical depth is  
evaluated by the relation 
\begin{equation} 
\tau_{\nu} = \kappa_{\nu} \rho \lambda_{\rm J}, 
\end{equation} 
where $\kappa_{\nu}$ is the opacity at the center. 
 
We neglect dark matter gravity in computing the collapse timescale. 
Although it might be important for the low-density regime,  
at sufficiently high densities, where fragmentation is expected to occur, 
the baryons have already contracted significantly owing to energy dissipation  
and dominate the gravity.  Indeed, fragmentation is caused by the self-gravity of the baryons 
and the contribution of the dark matter does not alter the dynamics. 
 
We initialize the model with an arbitrary set of initial conditions for the
density and temperature of the gas. In particular, we assume an initial
number density of $n_{\rm H}=0.1{\rm cm^{-3}}$ and, at this time, 
the temperature is set to be $T=300$K from  
Bromm et al. (2002)'s result (see their Figure 10 c).  
Indeed, the temperature evolution converges immediately to a certain trajectory,  
and variations in the initial conditions affect only the short transient phase  
of the initial evolution (e.g., Omukai 2000). 
For this reason, in this paper we do not consider different initial conditions 
and we refer the interested reader to,  
e.g., Palla et al. (1983) and Omukai (2000) for a thorough discussion of this issue. 
The initial ionization degree and H$_2$ concentration are assumed to be  
the values for uniform matter in the post-recombination universe: 
$y(e)=10^{-4}$, $y(H_{2})=10^{-6}$. 
Initially, carbon is fully ionized and Helium and Oxygen are neutral. 
 
\section{Results} 
In this section, we first present the thermal and chemical evolution  
for different initial values of the gas metallicity (\S 3.1); then, 
a reduced chemical model for low-metallicity clouds is presented (\S 3.2).
Using the derived effective equation of state, we discuss  
the fragmentation properties of prestellar clouds and estimate  
the typical mass-scales of fragmentation in Appendix.

\subsection{Thermal and Chemical Evolution} 
\subsubsection{Results for the Fiducial Cases} 
The temperature evolution for prestellar clouds for different metallicities  
is shown in Figure \ref{fig:nT.Tr3}. 
The cases with metallicities $[Z/H]= -\infty$ (i.e., Z=0  metal-free case),  
-5, -3 and -1 ($[Z/H]=-6, -4, -2$ and 0) are indicated by solid (dashed) lines. 
As an external radiation field, only the present-day (2.73K) CMB
is included although its effect can be neglected except for the lowermost temperature  
regime in the [Z/H]=0 case. We refer to this set of models as the fiducial cases hereafter. 
The dotted lines indicate those of constant Jeans mass. 
The dashed line labeled ``$\tau_{J}=1$'' shows the location where  
the central part of a cloud becomes optically thick to continuum  
radiation (eq. \ref{eq:univadiab}; see discussion later). 
Before this condition is met, i.e. to the left of the line,  clouds are still  
optically thin to continuum.    
 
The evolution of clouds with metallicity $[Z/H]=-(4-3)$ at  
$n_{\rm H} \sim 10^{4-8} {\rm cm^{-3}}$  
is mostly affected by the improvements of the model, namely by the 
inclusion of HD cooling and the modification of the collapse timescale. 
Conversely, both at lower and higher metallicity values,  
the thermal evolution is hardly altered. 
 
In Figure 2, we illustrate separately the contribution by each processes  
to cooling/ heating rates in the fiducial cases. 
For the same cases, we also show the evolution of H$_2$ and HD fractions in Figures 3 and 4,  
respectively. 
In terms of major coolants, the thermal evolution  
can be classified into the following three metallicity ranges: 
(i) quasi-primordial clouds ($[Z/H] \la -6$)  
(ii) metal-deficient clouds  ($-5 \la [Z/H] \la -3$) 
(iii) metal-enriched clouds  ($-2 \la [Z/H]$). 
For each of these ranges, we now describe the evolution presented in Figures 1-4. 
 
\begin{enumerate} 
 
\item Quasi-Primordial Clouds ($[Z/H] \la -6$) 
 
The presence of metals at metallicity levels as low as $[Z/H] \la -6$ does not 
significantly alter the thermal evolution in any density range.  
The evolution of both temperature and chemical species follows closely  
those of the metal-free case. 
Molecular hydrogen is always an important cooling agent.  
HD hardly affects the overall evolution despite contributing  
the cooling as much as H$_2$ at $\sim 10^{5} {\rm cm^{-3}}$  
(see also Bromm et al. 2002).  
 
To clarify the reasons for this trend, we first describe the HD formation process. 
The abundance of HD is determined by the equilibrium between the formation  
reaction (D4 reaction in Table 1) and its inverse dissociation reaction  
(D6),  
\begin{equation} 
D4, D6: {\rm D^+    +   H_2   \leftrightarrow  H^+  +    HD }. 
\end{equation} 
The deuterium ionization degree is set by the equilibrium between D1 and D2  
reactions,
\begin{equation} 
D1, D2: {\rm D    +   H^+   \leftrightarrow  D^+  +    H }. 
\end{equation} 
The HD to H$_2$ ratio is then  
\begin{equation} 
\frac{n({\rm HD})}{n({\rm H_2})}= \frac{k_{D4} n(D^{+})}{k_{D6} n(H^{+})} 
= \frac{k_{D4} k_{D1} n({\rm D})}{k_{D6} k_{D2} n({\rm H})} 
=2 {\rm exp}(\frac{421 {\rm K}}{T}) \frac{n(D)}{n(H)}, 
\label{eq:HDfract} 
\end{equation} 
where we have used the reaction coefficients given in Table 1. 
From equation (\ref{eq:HDfract}), we see that the HD is formed  
abruptly if the temperature is reduced significantly with respect to the value corresponding to the energy  
difference between H$_2$ and HD (421K). 
Specifically this occurs when the temperature falls below about 150K. 
 
In the case of quasi-primordial clouds, HD becomes an  
important coolant at $\sim 10^{5}{\rm cm^{-3}}$, after H$_2$ has
already reached LTE. Almost at the same time  
HD reaches LTE, cooling becomes inefficient,  and
the temperature increases with increasing density.   
As a result of temperature increase, HD is dissociated immediately 
(see Figure \ref{fig:fHD}). 
This is the reason why HD does not affect the thermal evolution  
in the lowest metallicity cases. 
  
H$_2$ is formed via the H$^{-}$ channel (combination of reactions H2 and H3)  
in the low density regime $\la 10^{8} {\rm cm^{-3}}$. 
Because of recombinations of electrons, the catalysts of this reaction, 
the H$_2$ fraction saturates.  
This results in the plateau in H$_2$ fraction $f_{\rm H_2} \sim 10^{-3}$  
in Figure \ref{fig:fH2}.  
Hydrogen becomes fully molecular only after the three-body reaction  
becomes important at densities higher than the threshold value  
$n_{\rm H} \sim 10^{8} {\rm cm^{-3}}$ (Palla et al. 1983). 
The rapid increase in H$_2$ fraction for $n_{\rm H} \ga 10^{8}  
{\rm cm^{-3}}$ is owing to the three-body reaction (Figure \ref{fig:fH2}). 
 
During the active phase of three-body H$_2$ formation, the heating is  
dominated by liberation of the chemical binding energy  
(Figure \ref{fig:cool} a,b). 
It exceeds the compressional heating by about an order of magnitude. 
After the hydrogen becomes almost fully molecular  
at $\sim 10^{12} {\rm cm^{-3}}$, the H$_2$ formation heating rate drops. 
However, almost simultaneously the H$_2$ line becomes optically thick  
and the cooling rate for a give temperature drops.  
Then the temperature continues to increase gradually.  
In the [Z/H]=-6 case, the grain thermal emission dominates a short interval  
around $10^{13-14} {\rm cm^{-3}}$ before dust evaporization at $\sim  
10^{14.5} {\rm cm^{-3}}$ (Figure \ref{fig:cool} b).   
Because of increasing temperature, the H$_2$ is about to dissociate  
at $\sim 10^{14} {\rm cm^{-3}}$.  
Indeed, a small fraction of H$_2$ dissociates.  
At this time, H$_2$ collision-induced emission (CIE) becomes important since  
the high density leads to frequent collisions and thus higher emissivity. 
CIE and its inverse collision-induced absorption are emission and absorption,
respectively, of photons from short-lived ``supermolecular'' species 
(i.e., ${\rm H_{2}-H_{2}, H_{2}-He}$, or ${\rm H_{2}-H}$) formed during 
collisions between H$_2$ and other particles. 
These supermolecules can emit and absorb photons with much 
higher probability than an isolated H$_2$ molecule because of their  
temporarily induced electric dipoles 
(e.g., Lenzuni, Chernoff, \& Salpeter 1991).  
Collision-induced emission dominates the cooling until it becomes  
optically thick to the self-absorption at $\sim 10^{16} {\rm cm^{-3}}$.
During this interval, the H$_2$ that was partially dissociated is re-assembled 
in a fully molecular form.
The alternating behavior of H$_2$ formation cooling and dissociation 
heating in Figure \ref{fig:cool} (a),(b) reflects this history.

After the cloud becomes optically thick to H$_2$ collision-induced absorption,
radiative cooling rate decreases rapidly (see Figure \ref{fig:cool} a, b).
However, the temperature continues to increase only gradually owing to 
the effective cooling by H$_2$ dissociation.
Because the temperature is already very close to the dissociation temperature,
the dissociation starts just after the clouds becomes optically thick. 
The temperature begins to increase almost adiabatically 
when the dissociation is nearly completed at $\sim 10^{20}{\rm cm^{-3}}$.
Finally, with some more compression owing to inertia, a hydrostatic core, 
or in another word, a protostar is formed at $10^{22} {\rm cm^{-3}}$ with a
mass-scale of $\sim 10^{-3} M_{\sun}$ (Omukai \& Nishi 1998).  
This will eventually grows to a massive main sequence star owing 
to the subsequent accretion of ambient matter (Omukai \& Palla 2001, 2003; Tan \& McKee 2004).

\item Metal-deficient Clouds ($-5 \la [Z/H] \la -3$)

For prestellar clouds in this metallicity range, the effects of metals 
become more visible.
Although H$_2$-line emission still dominates the cooling in the 
low-density regime ($\la 10^{7} {\rm cm^{-3}}$ for [Z/H]=-5;
$\la 10^{4} {\rm cm^{-3}}$ for [Z/H]=-4;
$\la 10^{3} {\rm cm^{-3}}$ for [Z/H]=-3; see Figure 
\ref{fig:cool} c-e),
an enhanced H$_2$ fraction due to grain surface 
reaction (reaction H8; see Figure \ref{fig:fH2}) results in a 
small deviation in temperature evolution from the metal-free case 
(Figure \ref{fig:cool}).
Below $\sim 10^{8} {\rm cm^{-3}}$, differences in the H$_2$ fractions at comparable densities 
between the metal-free case and other cases 
are a result of  grain-surface reaction.
Because of this reaction, the 
H$_2$ fraction continues to increase towards higher densities and 
the plateau in H$_2$ fraction $f_{\rm H_2} \sim 10^{-3}$ disappears.

The next evolutionary phase is dominated by HD cooling.
For $[Z/H]=-5$, this effect is not yet apparent, 
but for [Z/H]=-4 and -3, HD becomes the dominant coolant
in the range $\simeq 10^{4} - 10^{6} {\rm cm^{-3}}$ for [Z/H]=-4 
and $\simeq 10^{3} - 10^{5} {\rm cm^{-3}}$ for [Z/H]=-3 
(see Figure 4 c-e).
Unlike for quasi-primordial clouds, 
HD becomes important because additional H$_2$ formed 
by grain-surface reaction enhances cooling and
makes the temperature to fall below the critical value $\simeq 150$K for HD
formation.
Once this condition is satisfied, dramatic HD formation and 
resultant HD cooling further lowers the temperature.
HD formation is thus accelerated.
Also, unlike in the quasi-primordial clouds, 
the HD is not destroyed at densities $\ga 10^{5}{\rm cm^{-3}}$ 
because for  [Z/H]=-4 and -3 the temperature remains low
(Figure \ref{fig:fHD}).
 
At higher densities, the dominant cooling mechanism is hydroxyl and/or 
water line emission.
As shown in Figure \ref{fig:yO}, the major form of oxygen compounds depends on 
the temperature and thus on the metallicity.
Namely, for [Z/H]=-5 and -4, O is converted 
to H$_2$O through OH, while for [Z/H]=-3 a large part of 
O becomes O$_2$.
The relevant oxygen reactions are shown schematically in Figure \ref{fig:COchem} (a).
OH is mainly formed via the reaction
\begin{equation}
Z10: {\rm   O     +   H     \rightarrow   OH    +   \gamma}. 
\end{equation} 
For [Z/H]=-5, OH is immediately converted to H$_2$O 
by the reaction
\begin{equation}
Z12: {\rm   H_2    +   OH    \rightarrow    H_2O   +   H}
\end{equation}
at a density $\sim 10^{7} {\rm cm^{-3}}$.
For [Z/H]=-4, OH is formed at $\sim 10^{5} {\rm cm^{-3}}$
and remains in this form for some intervals because of the lower temperature 
with respect to the [Z/H]=-5 case. 
Some fraction of OH becomes O$_2$ via the reaction
\begin{equation}
Z19: {\rm   O     +   OH    \rightarrow   O_2    +   H}.
\end{equation}
The reaction coefficient of the H$_2$O-bearing reaction Z12 is
$k_{Z12} \propto T^{1.6}{\rm exp}(-1660K/T)$ while that of 
the O$_2$-bearing reaction Z19 is $k_{Z19} \propto T^{-0.5}{\rm exp}(-30K/T)$. 
Then, in a low temperature environment, H$_2$O formation is 
significantly suppressed and O$_2$ is the dominant oxygen compound with respect  to H$_2$O.
In particular, for $T \la 300$K, the reaction Z12 is frozen and 
O is locked in O$_2$ without forming H$_2$O.
At densities $\sim 10^{8}{\rm cm^{-3}}$, the temperature rises suddenly owing to 
heat injection by the three-body H$_2$ formation.
Consequently, OH and also O$_2$ become H$_2$O via the reactions 
Z20 and Z12, respectively.
For [Z/H]=-3, the temperature remains low ($\la 200$K) and
the majority of O becomes O$_2$ and remains in this form without changing 
to H$_2$O.

During this OH/H$_2$O cooling phase, H$_2$ formation heating exceeds the 
compressional heating (Figure 2 c-e).
As a consequence, the temperature increases at 
$\la 10^{8} {\rm cm^{-3}}$ for [Z/H]=-5,
($\la 10^{8} {\rm cm^{-3}}$ for  [Z/H]=-4, and 
$\la 10^{5} {\rm cm^{-3}}$ for [Z/H]=-3, respectively).
For [Z/H]=-5, the temperature is sufficiently high to hide this effect which
instead becomes apparent for [Z/H]=-4.
When [Z/H]=-4, heat injection due to the abrupt onset 
of the three-body reaction causes a dramatic temperature rise at 
$\sim 10^{8}{\rm cm^{-3}}$ since the temperature in the preceding evolution 
is low. 
On the other hand, for [Z/H]=-3, heat injection is more gradual 
since the majority of H$_2$ is formed by the more gradual grain-surface reaction.

When H$_2$ formation dominates the heating mechanism, 
the major cooling agent is represented by dust thermal emission 
(see Figure 4c-e).
Dust and gas exchange energy by collisions. 
At high density, frequent collisions lead to a higher cooling rate
($\Lambda \propto n$ per unit mass).
In spite of this cooling, the temperature continues to increase
as long as H$_2$ formation heating is important.

When H$_2$ formation is completed, the heating rate drops, and
the temperature begins to fall rapidly due to dust cooling.
Dust cooling remains dominant until the clouds
become optically thick to dust self-absorption, at 
densities in the range $10^{14-13} {\rm cm^{-3}}$ for $-5<[Z/H]<-3$.
After that, the cloud contracts adiabatically and temporary hydrostatic cores 
(first cores) might be formed as for present-day star formation 
(Larson 1969; see below).
When the temperature reaches 2000K, H$_2$ dissociation occurs.
The evolution after this point is exactly the same as that of the
quasi-primordial clouds.

\item Metal-enriched Clouds~($-2 \la [Z/H]$)

When metal-enrichment proceeds up to $[Z/H] \simeq -2$, 
the thermal evolution of prestellar clouds becomes 
qualitatively the same as the present-day one.
Series of coolants become dominant one after the other
in the same order.
Only the characteristic density of each phase is shifted 
towards lower densities at higher metallicities.
Note the similar shapes in the temperature-evolution curves 
in this metallicity range in Figure \ref{fig:nT.Tr3}.
The rapid initial temperature drop is induced by fine-structure line
emission of C$^{+}$, C and O.
Unlike in the metal-deficient clouds, HD cooling does not affect 
the thermal evolution since C, O cooling are dominant.
The temperature increase (at densities $10^{3-6} {\rm cm^{-3}}$, 
$10^{2-4.5} {\rm cm^{-3}}$ for [Z/H]=-2 and -1, respectively) 
and stagnation (at  $10^{0-3} {\rm cm^{-3}}$ for [Z/H]=0) 
are due to H$_2$ formation heating although the grain-surface 
reaction deposits only a few \% of the binding energy 
(0.2 eV of the total 4.48 eV in the lowest density regime); 
the rest of the energy is used for exciting H$_2$ ro-vibrational lines or deposited 
on the grain and  eventually radiated away in both cases 
(Hollenbach \& McKee 1979).

In the meantime, atomic carbon is converted to carbon monoxyde (CO). 
Major carbon-related reactions, including CO formation, are shown 
schematically in Figure \ref{fig:COchem} (b).
CO cooling is dominant in the density range
$10^{3.5}-10^{5.5}{\rm cm^{-3}}$ 
($10^{3}-10^{5}{\rm cm^{-3}}$) for [Z/H]=-2  ([Z/H]=-1) 
although for [Z/H]=0 it is significant only for a
short time interval  around $10^{3} {\rm cm^{-3}}$ (Figure 2 f-h).
In some other studies, CO can be the dominant coolant for a wider range 
of densities even for [Z/H]=0 (e.g., Koyama \& Inutsuka 2000).
Probably this is due to different assumptions on the external radiation 
field and on the effects of cosmic rays.
At higher densities, infrared thermal emission by dust grains becomes
important.
The gas temperature decreases until it reaches the dust temperature, 
and then thermally couples to dust grains.
This occurs at densities $\simeq 10^{8} {\rm cm^{-3}}$ for [Z/H]=-2, 
$\simeq 10^{6.5} {\rm cm^{-3}}$ for [Z/H]=-1, and 
$\simeq 10^{5} {\rm cm^{-3}}$ for [Z/H]=0.
After that, the thermal evolution is controlled by dust cooling 
and the temperature increases gradually with increasing density.
When the central parts of the clouds become optically 
thick to dust self-absorption, the temperature begins to increase 
adiabatically.
The evolution thereafter is the same as in the previous case.
\end{enumerate}

After the clouds become optically thick to continuum,
the temperature increases adiabatically. 
As seen in Figure \ref{fig:nT.Tr3}, all the paths in the adiabatic contraction 
converge approximately to the dashed line in spite of different  
values of temperature and density when the clouds become optically 
thick.
The reason for this convergence has already been explained in Paper I,
but we repeat it here briefly for later use. 
First, let us consider the temperature evolution paths just 
before the clouds become optically thick.
Since cooling is dominated by continuum emission 
(i.e., the H$_2$ collision-induced emission or the dust 
thermal emission depending on metallicity) and the heating is dominated
by compressional heating, the temperature is determined by 
the balance between them: 
\begin{equation}
4 \kappa_{Z} \sigma T^4 = \frac{c_{s}^2}{t_{\rm col}}, 
\label{eq:Tpath}
\end{equation}
where $\kappa_{Z}$ is the continuum opacity for gas with metallicities $Z$ 
and we use the pressure-free free-fall timescale 
$t_{\rm col}=t_{\rm col, 0}=\sqrt{3 \pi/32 G \rho}$ for this analytic estimate.
Note that the temperature evolution, $T=T(n_{\rm H})$, as given by 
equation (\ref{eq:Tpath}), depends on metallicity through the 
opacity $\kappa_{Z}$.
While the density increases along this path, 
the central region, whose length scale
is $\sim \lambda_{\rm J}$, becomes optically thick at the position in 
the $n_{\rm H}-T$ plane where the condition  
\begin{equation}
\tau_{J} \equiv \kappa_{Z} \lambda_{J}=1
\label{eq:tau1}
\end{equation}
is satisfied.
Eliminating $\kappa_{Z}$ from equations (\ref{eq:Tpath}) and (\ref{eq:tau1}), 
we obtain the relation
\begin{equation}
T=(\frac{k^3}{12 \sigma^2 m_{\rm H}})^{1/5} n_{\rm H}^{2/5}. 
\label{eq:univadiab}
\end{equation}
This is shown by the dashed line denoted as ``$\tau_{\rm J}=1$'' 
in Figure 1.
So far, the meaning of this line has been that of the line connecting 
all the points where the central parts of the clouds become optically 
thick for different metallicities.
However, this line has another significance.
Note that the exponent of $n_{\rm H}$ (=2/5) in this relation 
is accidentally equal to 
$\gamma_{\rm ad}-1 \simeq 0.4$ for molecular hydrogen.
Then, this line coincides with an adiabatic path 
$T \propto n_{\rm H}^{\gamma_{\rm ad}-1}$.
Since this path passes through the points where 
the clouds become optically thick, this is indeed the adiabatic 
evolutionary line in optically thick phase for any $Z$.
This is the reason for the convergence of the adiabatic paths.
In other words, when the clouds becomes optically thick, 
their specific entropy has the universal value.
The evolution in the optically thick phase follows just the 
constant entropy path with this universal adiabat, 
if entropy in chemical binding energy is taken into consideration.

\subsubsection{Effects of Cosmic Microwave Background Radiation}
Next we discuss the effect of an external radiation field on 
prestellar thermal evolution.
In this paper we consider only the CMB in high-redshift universe.
Figure 7 shows the cases irradiated by the black-body radiation with 
$T_{\rm rad}= 54.6K$ corresponding to $1+z=20$, the typical epoch for 
the first star formation in the universe.
The cases without strong CMB (fiducial cases shown in Figure 1) are 
indicated by dotted lines for comparison.

As expected, the temperature cannot drop below $T_{\rm rad}$.
On the other hand, the behavior at higher temperatures is hardly affected 
by the CMB radiation.  
An exception to this is the case of [Z/H]=-3 around 
$\sim 10^{8} {\rm cm^{-3}}$, where the temperature jumps up
owing to heat injection by three-body H$_2$ formation in the strong 
CMB case.
In this case, as a result of suppression of 
H$_2$ formation on grains by higher dust temperature,
three-body reaction becomes more important than in the fiducial case.

Another difference is that the temperature in the adiabatic contraction 
phase does not coincide with the universal adiabat. 
The reason is clear from the discussion above, where 
we have assumed that heating is dominated 
by compressional heating when the clouds becomes optically thick.
For higher metallicities, this condition is violated 
because of  CMB heating of dust grains. 
Despite the non-convergence in the adiabatic phase, 
after this phase these evolutionary paths converge to 
that of the zero-metallicity case. Indeed, 
the evaporation of dust grains suddenly decreases the optical depth 
in the central parts of the clouds.
After that, the clouds evolve by  H$_2$ CIE cooling, 
exactly in the same as in the zero-metallicity case. 
This is partly an artifact of our one-zone treatment 
because in the envelope we expect dust grains to survive and to remain optically thick
even though the central parts has become optically thin again.
In reality, photons emitted in the central region are absorbed in the 
envelope that is heated and that begins to back-warm the central 
region.
If this effect is important, our treatment breaks down.
To explore this, a radiation hydrodynamic study in necessary is required.
This is beyond the scope of this paper, and we postpone it for a future study.

\subsubsection{Role of Metals in Gas and Dust Phases: A Minimum Model for
Metallicity Effects}
Metals can be either in gas phase or dust grains.
These two forms of metals influence the thermal evolution in different ways.
Grains are of importance to    
(i) cooling by dust grains, and 
(ii) H$_2$ formation on grain surface,
while gas-phase metals act via 
(iii) cooling by atomic/molecular metal lines.
To see these effects separately, we show the same cases 
as the fiducial ones (Figure \ref{fig:nT.Tr3}) 
but without gas-phase metals in Figure \ref{fig:nT.dustonly}, and
without dust grains in Figure \ref{fig:nT.nodust}. 
Other parameters are the same as in the fiducial cases, whose temperature
evolution are shown by dotted lines for comparison.
Although we still use the same denomination [Z/H] for easier 
comparison with the fiducial cases, the amount of total metals are 
smaller than that in the fiducial cases with the same [Z/H] 
because one form of metals is missing.

Figure \ref{fig:nT.dustonly} shows that with dust grains only, 
the thermal evolution for gas metallicities [Z/H] $\la -3$ is approximately 
reproduced.
For higher gas metallicities ([Z/H] $\ga -2$), the behavior in the low 
density regime is strongly altered because of the lack of the fine-structure-line 
cooling.
In the metallicity range [Z/H]$\simeq$ -(5--3), cooling in the intermediate 
density regime is dominated by OH and H$_2$O, which have not been included in this 
Figure.
Despite this, there is only a rather small deviation in temperature 
since cooling by H$_2$ and HD have comparable effects.
Of the two effects of dust grains, dust cooling becomes important
at lower metallicity than the effect on H$_2$ formation. 

Figure \ref{fig:nT.nodust} shows the predicted thermal evolution when no dust 
is present. With gas-phase metals only, the temperature behavior is completely 
altered.
From the comparison between Figures \ref{fig:nT.dustonly} and \ref{fig:nT.nodust},
we can conclude that at low metallicities cooling by dust 
grains is dominant and the thermal evolution can be reproduced even without 
gas-phase metals.
On the other hand, in metal-enriched cases, cooling by fine-structure 
lines are important in the low density regime.

Considering only the dust and the fine-structure cooling,
we can construct a minimum model of metallicity effects.
In addition to the metal-free model which includes H$_2$ and HD cooling, 
H$_2$ collision-induced emission/absorption, and H1-8, D-1-6 reactions,
we include fine-structure line cooling and recombination of C$^{+}$ (Z28).
The result is shown in Figure \ref{fig:nT.smchem}.
Indeed, this model provides a good approximation of the thermal evolution.
Although there are still small deviations owing to the lack of 
OH and H$_2$O cooling, this model significantly simplifies the
implementation of metallicity-dependent effect, since it
does not solve the complex metal chemistry nor 
the level populations of C and O molecules. 
As the fine-structure lines are two (CII) or three (CI, OI) level systems,
we can easily include them.

\subsection{Reduced Chemical Model for Low-metallicity Clouds}
Although the minimum model above succeeds in roughly reproducing 
the thermal evolution of our full model, there are still some deviations.
As a further step, here we develop a model that reproduces 
the temperature evolution as well as the abundances of the dominant chemical 
species.

In the full model, we have incorporated about 500 reactions
(478 reactions in paper I and 18 deuterium reactions) 
between 50 species 
(${\rm H}$, ${\rm H_2}$, ${\rm e^-}$, ${\rm H^+}$, ${\rm H_2^+}$, 
${\rm H_3^+}$, ${\rm H^-}$, ${\rm He}$, ${\rm He^+}$, ${\rm He^{++}}$, 
${\rm HeH^{+}}$, 
${\rm D}$, ${\rm D^{+}}$, ${\rm D^{-}}$, ${\rm HD}$, ${\rm HD^{+}}$, 
${\rm C}$, ${\rm C_2}$, ${\rm CH}$, ${\rm CH_2}$, 
${\rm CH_3}$, ${\rm CH_4}$, ${\rm C^+}$, ${\rm C_2^+}$, ${\rm CH^+}$, 
${\rm CH_2^+}$, ${\rm CH_3^+}$, ${\rm CH_4^+}$, ${\rm CH_5^+}$, ${\rm O}$, 
${\rm O_2}$, ${\rm OH}$, ${\rm CO}$, ${\rm H_2O}$, ${\rm HCO}$, 
${\rm O_2H}$, ${\rm CO_2}$, ${\rm H_2CO}$, ${\rm H_2O_2}$, ${\rm O^+}$, 
${\rm O_2^+}$, ${\rm OH^+}$, ${\rm CO^+}$, ${\rm H_2O^+}$, ${\rm HCO^+}$, 
${\rm O_2H^+}$, ${\rm H_3O^+}$, ${\rm H_2CO^+}$, ${\rm HCO_2^+}$ and
${\rm H_3CO^+}$).
Apparently, not all of these reactions are of equal importance.
In many cases, we are only interested in reproducing the thermal history 
accurately and we are not interested in reproducing the evolution of all 
chemical species correctly.
To this end, we need to properly account for the abundances of the dominant 
coolants.
We select a reaction network which reproduces the abundance of 
dominant coolants, as well as dominant H, C and O compounds and the electron 
fraction in any evolutionary stage before the prestellar clouds becomes optically thick. 
This model includes 23 species 
${\rm H}$, ${\rm H^+}$, ${\rm H^-}$, ${\rm H_2}$, ${\rm e}$,
${\rm D}$, ${\rm D^+}$, ${\rm HD}$, ${\rm O}$, ${\rm OH}$, ${\rm H_2O}$, 
${\rm O_2}$, ${\rm O^{+}}$, ${\rm OH^{+}}$, ${\rm H_2O^{+}}$, 
${\rm H_3O^{+}}$, ${\rm O_2^{+}}$, ${\rm C^+}$, ${\rm C}$, ${\rm CH}$, 
${\rm CH_2}$, ${\rm CH_3}$, ${\rm CH_4}$, ${\rm CO}$, ${\rm CO^+}$, 
${\rm CO_2}$ and 54 reactions, which are given in Table 1.
Reactions related to C and O 
are shown schematically in Figure \ref{fig:COchem}. 
 
In Figure \ref{fig:nT.minichem}, we present the temperature evolution predicted
by the model with full network and that by the reduced one. 
The agreement is so good that the curves appear to be degenerate in the plot.
The corresponding evolution of the dominant chemical 
species are also shown in Figure \ref{fig:ymini}, for comparison. 
Although there are still some deviations when the fractional abundances 
are low, they are not important coolants at those times.
When some species are either important coolants or represent the dominant 
compound of a given element, their abundances are 
well reproduced by the reduced model.

\section{Summary and Discussion}
In this paper, we have studied the thermal and chemical evolution 
of star-forming clouds with different metallicities. 
We assume that the collapse rate of the clouds is  
approximately at the free-fall rate and the characteristic radius 
of collapsing clouds is given by the Jeans length.
The model developed in Omukai (2000) has been updated under some aspects, 
most notably by inclusion of deuterium chemistry.
The temperature evolution obtained from this model has been presented 
in Figure \ref{fig:nT.Tr3}.
The HD cooling modifies evolution of the clouds in the metallicity range 
$Z \sim 10^{-(3-5)}Z_{\sun}$, this species being the dominant coolant in the 
density range $\sim 10^{4-6}{\rm cm^{-3}}$.

While we have solved the thermo-chemical microphysics in great detail, 
we have adopted a simplified dynamical evolution.
Our full thermal modeling is numerically too costly to be coupled
to detailed hydrodynamical models.
To favor such experiments,  though, a reduced model retaining 
the essential features of the full model is required.
As a simplest model of this kind, we have shown that a model including 
only fine-structure line cooling and dust process, i.e. dust-cooling and 
H$_2$ formation on dust surface, in addition to the metal-free gas physics, 
reproduces the temperature evolution very well.
A better (but more time-consuming) model has also been implemented 
with reduced chemistry (with about 20 species 
and 50 reactions) which can reproduce the thermal evolution 
as well as the chemical evolution of important species.

In Appendix, by applying the linear theory of cloud cores deformation to 
the above temperature evolution, we have discussed 
fragmentation history of the star-forming clouds.
Given the initial distribution of elongation, 
the mass spectrum of fragments can straightforwardly be derived from this theory.
As examples, we have shown the mass spectrum for two 
different initial distributions of core elongations:
(i) a uniform initial distribution,
and 
(ii) a Gaussian distribution peaking around half of the critical value 
of fragmentation with variance of 0.1.
In both cases, we have found that while in the metal-free case 
fragmentation is possible only at the end of the efficient 
molecular cooling phase, 
fragmentation at high density $\la 10^{10}{\rm cm^{-3}}$ is enabled
even with a slight ($[Z/H] \sim -6$) enrichment level via dust grains,
the probability of this fragmentation increases drastically 
around $[Z/H] \sim -5$.
Since the fragmentation mass-scale is determined by the thermal Jeans scale 
at the fragmentation epoch, this results in sub-solar mass fragments 
even for slightly metal-enriched cases.

According to observations of nearby star-forming regions, 
stellar initial mass function (IMF) does not appear to depend strongly 
on metallicity (e.g., Kroupa 2002).
However, the metallicity range studied is still very limited (a factor of a few). We predict that if systems with metallicities around $[Z/H] 
\sim -4$ could be sampled, their IMF should resemble that from the present results.
In the quasi-primordial case $[Z/H] \la -6$, very massive fragments 
are predominantly produced.
Even in this case if extremely strong UV field 
is present photodissociation might lead to low-mass star formation 
by atomic cooling (Omukai \& Yoshii 2003), although 
Bromm \& Loeb (2003b) argue an alternative possibility that a supermassive 
object is the likely outcome for the atomic-cooling cloud.
For modestly metal-enriched ($[Z/H]\la -1$) clouds, fragmentation 
can occur at two different stages during the collapse, i.e. both in the molecular-cooling phase at lower density 
and in the dust-cooling phase at higher density.
The former fragmentation produces predominantly massive stars, 
although the typical mass-scale decreases with increasing metallicity;
the latter produces sub-solar mass stars.
The relative importance of these two fragmentation modes changes with 
metallicity.
Dust-induced fragmentation is of increasing importance for 
clouds with higher metallicity. Reflecting the two modes of fragmentation, 
the IMF often shows two peaks in low-metallicity star-forming clouds.
As metal enrichment increases to the level of $[Z/H] \sim -1$, these two peaks 
eventually merge as the line- and dust-cooling phases become indistinguishable 
in temperature evolution. This might explain why the present-day IMF is  single-peaked.

In this work, we have assumed the same dust-size distribution 
as the standard MRN distribution for local ISM 
(Mathis, Rumpl, \& Nordsieck 1977).
The origin of dust grains in the local ISM is largely attributed 
to asymptotic giant branch (AGB) stars with smaller contribution from 
Type II SNe. In the early universe  dust mainly comes from Type II (Todini \& Ferrara 2001) 
or pair-instability SNe (Nozawa et al. 2003; Schneider, Ferrara \& Salvaterra 2004), since there could not 
be sufficient time for star to evolve to the AGB phase.
The average SN grain radius is smaller than for AGB stars, and thus the total area per unit mass of dust 
is larger than the standard MRN distribution used in this work.
The large grain area enhances both dust-gas thermal coupling and the H$_2$ formation rate on grain surface.
Therefore, even small dust amounts can produce pronounced effects on the chemistry.  
Also the elemental abundance ratios in ejecta from the first SNe are different
from those in local ISM, which we have used in the analysis in this paper. 
Elements important for the cooling are carbon and oxygen.
In comparison with the local ISM, carbon is significantly deficient 
($<1/100$) in the first SNe ejecta while the amount of oxygen is similar in 
both environments.
Among these two elements, in low metallicity environment
atomic and molecular forms of oxygen are more important coolants 
than those of carbon (\S 3).
As gas-phase oxygen abundance is similar,  
difference in gas-phase abundance ratio does not alter thermal evolution 
of low metallicity clouds significantly. 
We plan to study effects of different grain size distribution and gas phase 
elemental abundances in a forthcoming paper.

Recently, Bromm \& Loeb (2003a, BL03) have also discussed the critical 
metallicity separating low- and high-mass star formation modes.
In their analysis, they included only fine-structure-line cooling of 
C$^{+}$ and O, and derived the critical metallicity by 
estimating the amount of metals above which the fine-structure-line
cooling becomes more efficient than the compressional heating (at density 
$\simeq 10^{4} {\rm cm^{-3}}$) where the temperature drops to a minimum 
($\simeq 200$K) and fragmentation is expected to occur.
For cases in which either [CII] or [OI] is present, 
they obtained critical metallicity values of $[C/H]_{\rm crit}= -3.5$ and  
$[O/H]_{\rm crit}=-3.05$, respectively, by comparing the cooling rates at 
$10^{4} {\rm cm^{-3}}$ and 200K.
The comparison between their results and those derived in this work is shown in Figure \ref{fig:nT.BL},
where we plot the temperature evolution calculated by including only (a) [CII] or  (b) [OI] 
cooling in addition to the metal-free thermo-chemical processes.
In the case  (a) in which only [CII] cooling is included, the position of 
temperature minimum is consistent with BL03 analysis. 
For the (b) [OI] case, however, despite BL03 claim that $[O/H]_{\rm crit}=-3$, 
even with $[O/H]=-4$ the position of the temperature minimum 
is already shifted to $n_{\rm H} \sim 10^{6}{\rm cm^{-3}}$. This is because [OI] cooling 
becomes  important only at densities higher than the value $ 10^{4}{\rm cm^{-3}}$ adopted by BL03.
Note that the temperature at $n_{\rm H} \sim 10^{4}{\rm cm^{-3}}$ is hardly affected.
In our model, in which dust grains as well as gas-phase metals are considered, 
at $[Z/H]=-4$, the temperature becomes much lower due to the combined effects of  grain-surface 
H$_2$ formation and the HD cooling. Thus, in the absence of dust cooling, 
the fragmentation mass-scale remains relatively high, $\ga 10M_{\sun}$, 
and sub-solar mass stars are not formed. Hence we conclude that the dust-induced fragmentation 
leads to low-mass star formation for metallicity values well below those predicted by BL03.

\acknowledgements 
We thank the referee, Volker Bromm, for comments that improved 
the presentation of the paper.
This study is supported in part by the Grants-in-Aid
by the Ministry of Education, Science and Culture of Japan
(16204012:KO,14740129:TT).

\appendix
\section{Fragmentation Mass-scale}
In this section, we present a semi-analytic description of the hydrodynamical 
evolution for prestellar clouds whose thermal evolution has been described above.
In particular, we discuss the fragmentation history allowing to estimate 
the characteristic mass-scale of the fragments.

Following the results of numerical simulations of the  collapse of metal-free prestellar
clouds (e.g., Bromm et al. 1999; 2002), we expect that these clouds 
experience fragmentation at the first temperature minimum, 
i.e. when the effective adiabatic exponent 
${\rm dlog} p/{\rm dlog} \rho \equiv \gamma$ exceeds 1 for the first time 
after the rapid cooling phase.
For  metal-free clouds, this happens at number densities 
$n_{\rm H} \sim 10^{4}{\rm cm^{-3}}$.
After this fragmentation, the clouds start the run-away collapse, 
which can be described by the Larson-Penston-type self-similar solution, which was 
originally found by 
Larson (1969) and Penston (1969) for the isothermal case, 
$\gamma=1$, and was extended for other values of $\gamma < 4/3$ by Yahil 
(1983). 
We assume that collapsing prestellar clouds follow 
the Larson-Penston-type self-similar solution with the effective 
adiabatic exponent $\gamma$ as derived from the thermal evolution 
discussed above.

In particular, we discuss whether run-away-collapsing clouds 
fragment again or not.
Linear theory on non-spherical deformation of the self-similar solution
has been developed by Hanawa \& Matsumoto (2000) and Lai (2000).
According to these authors, the amplitude of a perturbation grows in  
a power-low of density; if we define the elongation as ${\cal E} \equiv (b-a)/a$
where $a,b$ are minor and major axis lengths of the cores, then 
${\cal E} \propto \rho^{n}$ with $n$ being a constant for a given $\gamma$.
The power of growth $n$ is presented in Figure \ref{fig:linear} 
as a function of $\gamma$.
As Hanawa \& Matsumoto (2000) and Lai (2000) considered only values of the adiabatic exponent
in the range $0.9 < \gamma < 1.25$; outside  this range of values, we extrapolate 
$n$ in the following way.
For $\gamma > 1.25$, the elongation decreases rapidly with increasing density and
 we linearly extrapolate the power of growth $n$ from lower values of $\gamma$.
When $\gamma$ is reduced below 0.9, the solution approaches 
the homogeneous density distribution with the density contrast 
between the center and the envelope decreasing.
Eventually, at $\gamma=5/6$, the solution becomes 
the uniform-density one, for which the growth rate of 
elongation is $n=0.354$.
Here, we use the linear interpolation for $5/6 < \gamma < 0.9$
and for $\gamma \leq 5/6$ we use that for the uniform-density solution.
Our treatment of assuming the self-similar solution breaks down when 
$\gamma > 4/3$, where no self-similar solution exists.
In reality, when $\gamma$ exceeds $4/3$ at some density, 
a hydrostatic core forms at the center and 
it contracts owing to the accretion of the envelope after that.
This phase is treated by the same prescription as in \S 2.

When the elongation grows to a certain level of non-linearity,
fragmentation is expected to occur since filamentary clouds are 
gravitationally unstable (e.g. Larson 1985).
Given the lack of systematic investigation of the non-linear evolution, 
we use a simple prescription, neglecting complicated effects 
of, e.g., rotation etc.
Initially, just after the first fragmentation, 
prestellar clouds are assumed to be elongated with 
amplitude ${\cal E}_{\rm i} (\la 1)$.
At this point, a functional form for the probability distribution for ${\cal E}_{\rm i}$ is assumed.
As the clouds collapse, the elongations grow (or decay) 
following the linear theory.
When the elongation exceeds a threshold value 
${\cal E}_{\rm NL} \sim 1$, the cores becomes filamentary and 
those filaments fragment into two cores.
The mass of fragments are given by the thermal Jeans mass at the 
fragmentation.
Although the cores might grow further by accretion of the surrounding 
gas after fragmentation, this effect is not included.
The elongation of newly produced fragments obeys
the same probability distribution as that of the initial cores. 
Evolution of those fragments thereafter is treated in the same way 
as the initial cores.
By repeating the above procedure, we can obtain a mass distribution of
fragments.
Since the probability distribution of elongation is unknown,  
we make two different assumptions:
(i) a uniform initial distribution of elongations
and 
(ii) a distribution peaking around half of the critical value 
of fragmentation with variance $\sigma=0.1$.
For each case, we have carried out Monte-Carlo simulations with 
sample number 8192. 
By reducing the sample number to half this value, we have confirmed 
that this is sufficient for convergence.

In Figure \ref{fig:elong}, we show the evolution of elongation for 
the fiducial cases.
If the elongation exceeds the threshold ${\cal E}_{\rm NL}$ 
at some point, the clouds are assumed to fragment and the elongation is reset 
to an initial value.
This effect is not included in this Figure.
Suppose this occurs at density $n_{\rm H, frag}$ and  
a new fragment is assigned with a value of elongation ${\cal E}_{\rm i,frag}$.
At higher density $n_{\rm H}>n_{\rm H, frag}$, 
the ratio ${\cal E}/{\cal E}_{\rm i,frag}$  
is given by $({\cal E}/{\cal E}_{\rm i})|_{n_{\rm H}}
/({\cal E}/{\cal E}_{\rm i})|_{n_{\rm H, frag}}$, using 
${\cal E}/{\cal E}_{\rm i}$ shown in the Figure.

In Figure \ref{fig:fragmass}, the mass distribution of fragments 
is shown by histogram for each metallicity case.
The two panels correspond to the different initial distributions of elongation.
The results in the two cases are qualitatively very similar.
As our model is still rather crude, we should not take these 
quantitative differences seriously and should focus only on 
qualitative agreement.
In the metal-free case, only massive ($\sim 10^{3}M_{\sun}$) fragments 
are produced without fragmenting in the run-away collapse phase.
This can be understood from Figure \ref{fig:linear}.
The maximum value of elongation is attained around 
$n_{\rm H} \sim 10^{16} {\rm cm^{-3}}$, but it is only as large as the
initial value ${\cal E}_{\rm i}$ and then less than ${\cal E}_{\rm NL}$.
In contrast, for metallicities as low as [Z/H]=-6, 
low-mass fragmentation is possible although the fraction in such objects 
is still small.
The reason is that since the maximum elongation exceeds the initial value, 
the core can fragment provided that the initial elongation is close to 
the critical value ${\cal E}_{\rm NL}$.
Since highly elongated cores are more numerous for the uniform 
${\cal E}_{\rm i}$ distribution, the fraction of low-mass fragments is 
higher in this case.   
When metallicity reaches [Z/H]=-5, the low-mass stars 
($\sim 0.1M_{\sun}$) outnumber very massive objects 
($\sim 10^{3}M_{\sun}$) because the maximum elongation 
(around $10^{15} {\rm cm^{-3}}$) is sufficiently higher than the 
initial value and then the elongation can grow sufficiently around 
$10^{15} {\rm cm^{-3}}$.
In other words, the temperature minimum by the dust-cooling becomes 
sufficiently deep and fragmentation at this point occurs with 
high probability. 
In the [Z/H]=-4 case, as a result of the sudden heat-up around 
$10^{8}{\rm cm^{-3}}$, 
elongation of cores decreases rapidly (see Figure \ref{fig:elong}).
Without seed elongation, the fragmentation does not occur 
at higher density even with dust cooling.
Therefore, fragmentation occurs only 
at the line-cooling phase and produces only rather massive 
$\sim 10M_{\sun}$ fragments.
Heat injection being gradual, the situation in the case of 
[Z/H]=-3 is rather similar to that of [Z/H]=-5.
Because fragmentation occurs at $10^{10}-10^{13}{\rm cm^{-3}}$ 
with high probability (see Figure \ref{fig:elong}), 
only a single peak of low-mass objects appears in the mass distribution.
For [Z/H]=-(2--1), two peaks appear in the mass distribution 
corresponding to two minima of the temperature evolution (see Figure 
\ref{fig:nT.Tr3}):
one is around $10^{2}M_{\sun}$, resulting from line-induced 
fragmentation around $10^{2-3}{\rm cm^{-3}}$, and the other 
around $1 M_{\sun}$ resulting from the dust cooling around $10^{6-7}{\rm cm^{-3}}$.
With [Z/H]=0, only a single peak appears $\sim 1 M_{\sun}$ since the 
two temperature minima almost merge into a single one 
(Figure \ref{fig:nT.Tr3}).
In metal-enriched cases, because of the lower temperature, 
the thermal evolution is more vulnerable to external heating effects
than in lower metallicity cases.
External heating by stellar/active-galaxy radiation or cosmic rays
is quite uncertain in primeval galaxies and it is not included in our 
calculation.
Because of this, results concerning the high-metallicity cases 
are to be taken with some care. 
With this caveat in mind, 
resemblance between our fragment mass 
distribution and the observed IMF is encouraging.
    
Our simple analysis did not take into account interaction between fragments
and accretion after fragmentation. 
However, it is known that the IMF in present-day star formation is 
largely shaped by 
the fact that stars form in clusters out of gas clouds that are pervaded 
by supersonic turbulence, through the complex dynamics of many interacting
and accreting protostars (e.g. Bate, Bonnel, \& Bromm 2003).
Such a clustered nature is also vital to predict the shapes of prestellar 
clouds themselves due to the gravitational interactions in the cluster.
Besides, fragmentation occurs not only in dense molecular cores, but also 
in massive circumstellar disks after formation of a central star 
(Bate et al. 2003).
Because of those processes not included in our analysis, fragment mass 
function obtained here should be regarded as showing only typical 
mass-scale rather than detailed mass distribution.
In order to study the core mass function more quantitatively in future, 
we need to consider all those cluster effects above.

\newpage


\newpage
\bigskip

\begin{deluxetable}{cllll}
\tablecaption{Reduced Chemical Reactions for Low-Metallicity Prestellar Clouds
\label{tab:react} }
\tablehead{
\colhead{Number} &
\colhead{Reaction} & 
\colhead{Rate Coefficient \tablenotemark{a}} & 
\colhead{Reference \tablenotemark{b}} & 
} 
\startdata
H1& ${\rm H^{+}    +   e     \rightarrow   H     +   \gamma}$
& $k_{H1}={\rm exp}[-28.6130338-0.72411256({\rm ln}(T({\rm eV})))$
&  A97 \nl
&
& ~~~~~~~~~~~~~~~$-2.02604473 \times 10^{-2}({\rm ln}(T({\rm eV})))^{2} $
&   \nl
&
& ~~~~~~~~~~~~~~~$-2.38086188 \times 10^{-3}({\rm ln}(T({\rm eV})))^{3} $
&   \nl
&
& ~~~~~~~~~~~~~~~$-3.21260521 \times 10^{-4}({\rm ln}(T({\rm eV})))^{4} $
&   \nl
&
& ~~~~~~~~~~~~~~~$-1.42150291 \times 10^{-5}({\rm ln}(T({\rm eV})))^{5} $
&   \nl
&
& ~~~~~~~~~~~~~~~$+4.98910892 \times 10^{-6}({\rm ln}(T({\rm eV})))^{6} $
&   \nl
&
& ~~~~~~~~~~~~~~~$+5.75561414 \times 10^{-7}({\rm ln}(T({\rm eV})))^{7} $
&   \nl
&
& ~~~~~~~~~~~~~~~$-1.85676704 \times 10^{-8}({\rm ln}(T({\rm eV})))^{8} $
&   \nl
&
& ~~~~~~~~~~~~~~~$-3.07113524 \times 10^{-9}({\rm ln}(T({\rm eV})))^{9}] $
&   \nl
H2& ${\rm H     +   e     \rightarrow   H^-    +   \gamma}$
&  $k_{H2}=1.4 \times 10^{-18} T^{0.928} {\rm exp}(- T/1.62 \times
  10^{4})$  
&    GP98 \nl
H3& ${\rm H^-    +   H     \rightarrow   H_2    +   e}$
& $k_{H3}=4.0 \times 10^{-9} T^{-0.17} ~~~~~~~~(T>300)$
&    GP98 \nl
&
& $k_{H3}=1.5 \times 10^{-9} ~~~~~~~~~~~~~~~~~(T<300)$
&   \nl
H4&  ${\rm H_2    +   H     \rightarrow 3 H}$
& $k_{H4} = k_{\rm H}^{1-a} k_{\rm L}^a$
&    \nl
&      
& ~~~~$k_{\rm L}=1.12 \times 10^{-10} {\rm exp}(-7.035 \times 10^{4}/T)$
&    SK87 \nl
&
& ~~~~$k_{\rm H}=6.5 \times 10^{-7} T^{-1/2}$
&    \nl
&
& ~~~~~~~~~~~${\rm exp}(-5.2 \times 10^{4}/T)[1-{\rm exp}(-6000/T)]$
&    \nl
&      
& ~~~~$a = (1+n/n_{\rm cr})^{-1}$ 
&    \nl
&
& ~~~~${\rm log}_{10}(n_{\rm
  cr}) = 4.0-0.416 {\rm log}_{10}(T/10^4)$
&  \nl
&
& ~~~~~~~~~~~~~~~~~~$-0.327 ({\rm log}_{10}(T/10^4))^2$ 
&   \nl
H5&  $ {\rm 3 H \rightarrow   H_2    +   H}$                
& $k_{H5}=5.5 \times 10^{-29} T^{-1}$
&    PSS83 \nl
H6&  ${\rm 2 H     +   H_2    \rightarrow 2 H_2}$                
& $k_{H6}=k_{H5}/8$
&    PSS83 \nl
H7&  ${\rm 2 H_2   \rightarrow 2 H     +   H_2}$
& $k_{H7} = k_{\rm high}^{1-a} k_{\rm low}^a$
&   SK87 \nl
&
& ~~~~$k_{\rm low} = 1.18 \times 10^{-10} {\rm exp}(-6.95 \times 10^{4}/T)$
&  PSS83 ($k_{\rm high}$)\nl
&
& ~~~~$k_{\rm high} = 8.125 \times 10^{-8}T^{-1/2}$
&   \nl
&
& ~~~~~~~~~~~~${\rm exp}(-5.2 \times 10^{4}/T) [1-{\rm exp}(-6.0 \times
10^{3}/T)]$ 
&    \nl
&
& ~~~~$a = (1+n/n_{\rm cr})^{-1}$ 
&    \nl
&
& ~~~~${\rm log}_{10}(n_{\rm
  cr}) = 4.845-1.3 {\rm log}_{10}(T/10^4)$
&  \nl
&
& ~~~~~~~~~~~~~~~~~~$+1.62 ({\rm log}_{10}(T/10^4))^2$ 
&   \nl
H8&  ${\rm 2 H     +   grain \rightarrow   H_2}$
& $k_{H8}=6.0 \times 10^{-17}(T/300 {\rm K})^{1/2} f_{\rm a}$
&    TH85 \nl
& 
& ~~~~~ $ \times [1+4.0 \times 10^{-2}(T+T_{\rm gr})^{1/2}$
&  \nl
&
& ~~~~~~~~~~~~~~~~~$+2.0 \times 10^{-3} T +8.0 \times 10^{-6} T^{2}]^{-1}$
& \nl
&
& ~~~~~~$\times Z/Z_{\rm local}$
& \nl
&
& ~~~~~~$f_{\rm a}=[1+ {\rm exp}(7.5 \times 10^{2}(1/75 -1/T_{\rm gr}))]^{-1}$
&   \nl
D1&  ${\rm D     +   H^+    \rightarrow   D^+     +   H }$
& $k_{D1}=3.7 \times 10^{-10} T^{0.28}{\rm exp}(-43/T)$
&    NU02 \nl
D2&  ${\rm D^+    +   H     \rightarrow   D     +    H^+ }$
& $k_{D2}=3.7 \times 10^{-10} T^{0.28} $
&    NU02 \nl
D3&  ${\rm D     +   H_2    \rightarrow  D     +   HD }$
& $k_{D3}=9.0 \times 10^{-11} {\rm exp}(-3876/T)$
&    NU02 \nl
D4&  ${\rm D^+    +   H_2   \rightarrow  H^+  +    HD }$
& $k_{D4}=2.1 \times 10^{-9} $
&    NU02 \nl
D5&  ${\rm HD     +   H    \rightarrow   H_2     +   D }$
& $k_{D5}=3.2 \times 10^{-11} {\rm exp}(-3624/T)$
&    NU02 \nl
D6&  ${\rm HD    +   H^+     \rightarrow H_2     +   D^+ }$
& $k_{D6}=1.0 \times 10^{-9} {\rm exp}(-464/T)$
&    NU02 \nl

Z1& ${\rm   O^+    +   H     \rightarrow   H^+    +   O}$
& $k_{Z1}=6.80 \times 10^{-10}$
&    M97 \nl
Z2&  ${\rm   H^+    +   O     \rightarrow    O^+    +   H}$
& $k_{Z2}=7.00 \times 10^{-10}{\rm exp}(-232/{\rm T})$
&    M97 \nl
Z3& ${\rm   O^+    +   H_2    \rightarrow   OH^+   +   H}$
& $k_{Z3}=1.70 \times 10^{-9}$
&    M97 \nl
Z4& ${\rm   OH^+   +   H_2    \rightarrow   H_2O^+  +   H}$
& $k_{Z4}=1.01 \times 10^{-9}$
&    M97 \nl
Z5& ${\rm   H_2O^+  +   H_2    \rightarrow   H_3O^+  +   H}$
& $k_{Z5}=8.30 \times 10^{-10}$
&    M97 \nl
Z6& ${\rm   H_2O^+  +   e     \rightarrow   O     +   H_2}$
& $k_{Z6}=2.00 \times 10^{-7}({\rm T}/300)^{-0.50}$
&    M97 \nl
Z7& ${\rm   H_2O^+  +   e     \rightarrow   OH    +   H}$
& $k_{Z7}=1.60 \times 10^{-7}({\rm T}/300)^{-0.50}$
&    M97 \nl
Z8& ${\rm   H_3O^+  +   e     \rightarrow   OH    + 2 H}$
& $k_{Z8}=6.50 \times 10^{-7}({\rm T}/300)^{-0.50}$
&    M97 \nl
Z9& ${\rm   H_3O^+  +   e     \rightarrow   H_2O   +   H}$
& $k_{Z9}=3.50 \times 10^{-7}({\rm T}/300)^{-0.50}$
&    M97 \nl
Z10& ${\rm   O     +   H     \rightarrow   OH    +   \gamma}$
& $k_{Z10}=9.90 \times 10^{-19}({\rm T}/300)^{-0.38}$
&    M97 \nl
Z11&  ${\rm   O     +   H_2    \rightarrow    OH    +   H}$
& $k_{Z11}=3.43 \times 10^{-13}({\rm T}/300)^{2.67}{\rm exp}(-3160/{\rm T})$
&    M97 \nl
Z12&  ${\rm   H_2    +   OH    \rightarrow    H_2O   +   H}$
& $k_{Z12}=1.55 \times 10^{-12}({\rm T}/300)^{1.60}{\rm exp}(-1660/{\rm T})$
&    M97 \nl
Z13& ${\rm 2 OH              \rightarrow   H_2O   +   O}$
& $k_{Z13}=1.65 \times 10^{-12}({\rm T}/300)^{1.14}{\rm exp}(-50/{\rm T})$
&    M97 \nl
Z14& ${\rm   H^+    +   OH    \rightarrow   OH^+   +   H}$
& $k_{Z14}=2.10 \times 10^{-9}$
&    M97 \nl
Z15& ${\rm   H^+    +   H_2O   \rightarrow   H_2O^+  +   H}$
& $k_{Z15}=6.90 \times 10^{-9}$
&    M97 \nl
Z16&  ${\rm H     +   OH    \rightarrow   H_2    +   O}$
& $k_{Z16}=7.00 \times 10^{-14}({\rm T}/300)^{2.80}{\rm exp}(-1950/{\rm T})$
&    M97 \nl
Z17&  ${\rm   H     +   H_2O   \rightarrow   OH    +   H_2}$
& $k_{Z17}=6.83 \times 10^{-12}({\rm T}/300)^{1.60}{\rm exp}(-9720/{\rm T})$
&    M97 \nl
Z18& ${\rm 2 O               \rightarrow   O_2    +   \gamma}$
& $k_{Z18}=4.90 \times 10^{-20}({\rm T}/300)^{1.58}$
&    M97 \nl
Z19& ${\rm   O     +   OH    \rightarrow   O_2    +   H}$
& $k_{Z19}=4.34 \times 10^{-11}({\rm T}/300)^{-0.50}{\rm exp}(-30/{\rm T})$
&    M97 \nl
Z20&  ${\rm   H     +   O_2    \rightarrow    OH    +   O}$
& $k_{Z20}=3.30 \times 10^{-10}{\rm exp}(-8460/{\rm T})$
&    M97 \nl
Z21& ${\rm   H^+    +   O_2    \rightarrow   O_2^+   +   H}$
& $k_{Z21}=2.00 \times 10^{-9}$
&    M97 \nl
Z22& ${\rm   O_2^+   +   e     \rightarrow 2 O}$
& $k_{Z22}=1.95 \times 10^{-7}({\rm T}/300)^{-0.70}$
&    M97 \nl
Z23& ${\rm   O     +   CH    \rightarrow   CO    +   H}$
& $k_{Z23}=6.60 \times 10^{-11}$
&    M97 \nl
Z24& ${\rm   C     +   OH    \rightarrow   CO    +   H}$
& $k_{Z24}=1.10 \times 10^{-10}({\rm T}/300)^{0.50}$
&    M97 \nl
Z25& ${\rm   C     +   O_2    \rightarrow   CO    +   O  }$  
& $k_{Z25}=3.30 \times 10^{-11}$
&    M97 \nl
Z26& ${\rm   C^+    +   O_2    \rightarrow   O^+    +   CO}$
& $k_{Z26}=6.20 \times 10^{-10}$
&    M97 \nl
Z27& ${\rm   OH    +   CO    \rightarrow   CO_2   +   H}$
& $k_{Z27}=1.00 \times 10^{-13}$
&    M97 \nl
Z28& ${\rm   C^+    +   e     \rightarrow   C     +   \gamma}$
& $k_{Z28}=4.40 \times 10^{-12}({\rm T}/300)^{-0.61}$
&    M97 \nl
Z29& ${\rm   C^+    +   OH    \rightarrow   CO^+   +   H}$
& $k_{Z29}=7.70 \times 10^{-10}$
&    M97 \nl
Z30& ${\rm   CO^+   +   H     \rightarrow   H^+    +   CO}$
& $k_{Z30}=7.50 \times 10^{-10}$
&    M97 \nl
Z31& ${\rm   C     +   H     \rightarrow   CH    +   \gamma}$
& $k_{Z31}=1.00 \times 10^{-17}$
&    M97 \nl
Z32&  ${\rm   C     +   H_2    \rightarrow    CH    +   H}$
& $k_{Z32}=6.64 \times 10^{-10}{\rm exp}(-11700/{\rm T})$
&    M97 \nl
Z33&  ${\rm H     +   CH    \rightarrow   C     +   H_2 }$
& $k_{Z33}=4.98 \times 10^{-11}$
&    M97 \nl
Z34& ${\rm   H_2    +   CH    \rightarrow    CH_2   +   H}$
& $k_{Z34}=2.38 \times 10^{-10}{\rm exp}(-1760/{\rm T})$
&    M97 \nl
Z35&   ${\rm H     +   CH_2   \rightarrow   CH    +   H_2 }$
&  $k_{Z35}=2.70 \times 10^{-10}$
&    M97 \nl
Z36& ${\rm   H_2    +   CH_2   \rightarrow    CH_3   +   H}$
&  $k_{Z36}=5.18 \times 10^{-11}({\rm T}/300)^{0.17}{\rm exp}(-6400/{\rm T})$
&    M97 \nl
Z37& ${\rm H     +   CH_3   \rightarrow   CH_2   +   H_2}$
&  $k_{Z37}=1.00 \times 10^{-10}{\rm exp}(-7600/{\rm T})$
&    M97 \nl
Z38& ${\rm   H_2    +   CH_3   \rightarrow    CH_4   +   H}$
&  $k_{Z38}=6.86 \times 10^{-14}({\rm T}/300)^{2.74}{\rm exp}(-4740/{\rm T})$
&    M97 \nl
Z39&  ${\rm H     +   CH_4   \rightarrow   H_2    +   CH_3}$
& $k_{Z39}=5.82 \times 10^{-13}({\rm T}/300)^{3.00}{\rm exp}(-4045/{\rm T})$
&    M97 \nl
Z40&  ${\rm   H_2 +   C     \rightarrow   CH_2    +   \gamma}$
&  $k_{Z40}=1.00 \times 10^{-17}$
&    M97 \nl
\enddata
\tablenotetext{a}{The temperature $T$ is in K, except otherwise noted.}
\tablenotetext{b}{A97; Abel et al. (1997), GP98; Galli \& Palla (1998),
  SK87; Shapiro \& Kang (1987), M97; Millar et al. (1997), 
  NU02; Nakamura \& Umemura (2002), PSS83; Palla et al. (1983), 
  TH85; Tielens \& Hollenbach (1985)}
\end{deluxetable}

\newpage
\centerline{\bf Figures}
 
\plotone{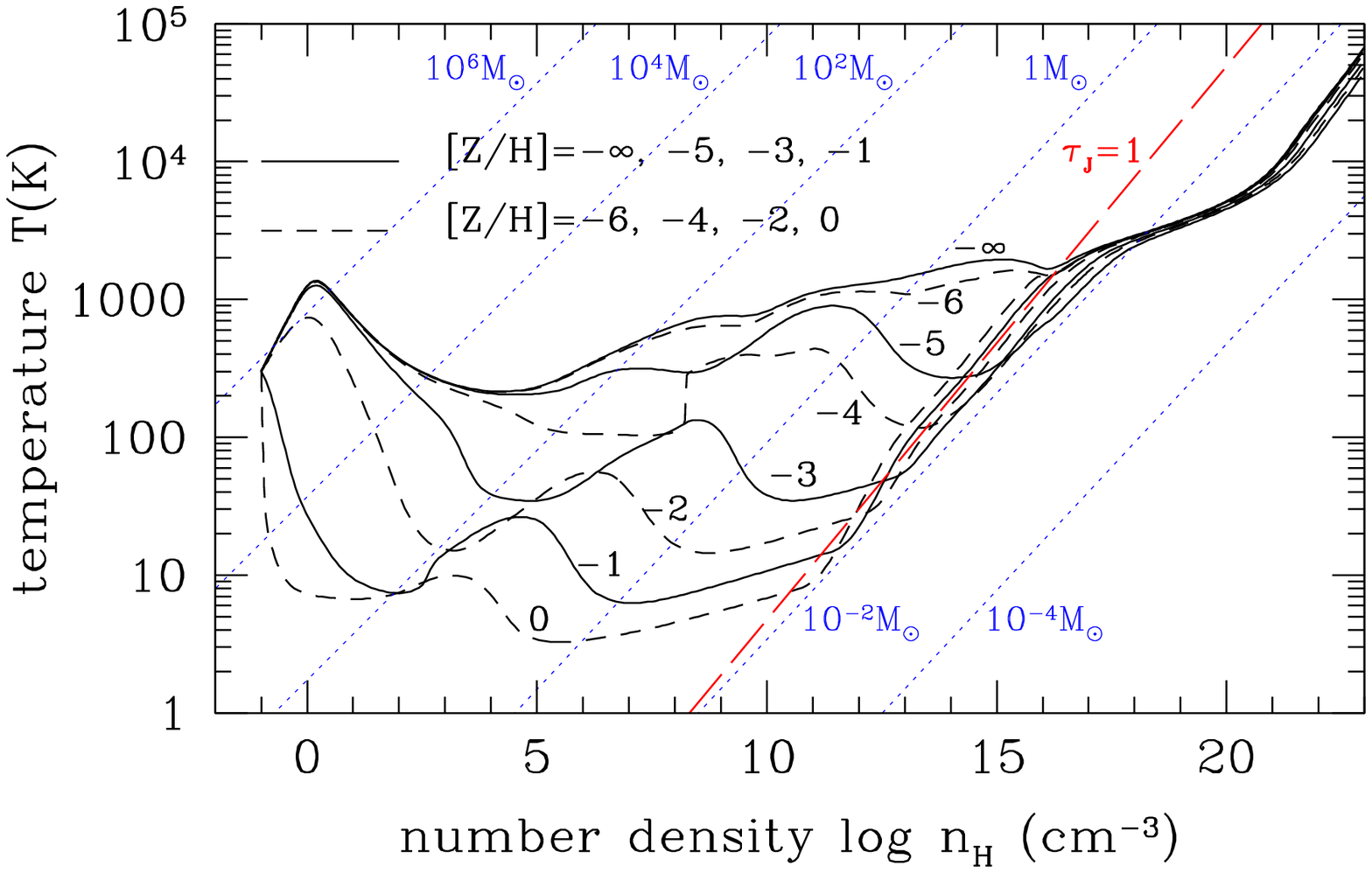}
\figcaption[nT.Tr3]{Temperature evolution of prestellar clouds
with different metallicities. Those with metallicities [Z/H]=-$\infty$ (Z=0), 
-5, -3, and -1 (-6, -4, -2, and 0) are shown by solid (dashed) lines.
Only the present-day CMB is considered as an external radiation field.
The lines for constant Jeans mass are indicated by thin dotted lines.
The positions where the central part of the clouds becomes optically thick 
to continuum self-absorption is indicated by the thin solid line 
(equation \ref{eq:univadiab}).
The intersection of the thin solid line with each evolutionary trajectory 
corresponds to the epoch when the cloud becomes optically thick 
to the continuum. 
To the right of this line, the clouds are optically thick and there is little 
radiative cooling.
\label{fig:nT.Tr3}}

\plottwo{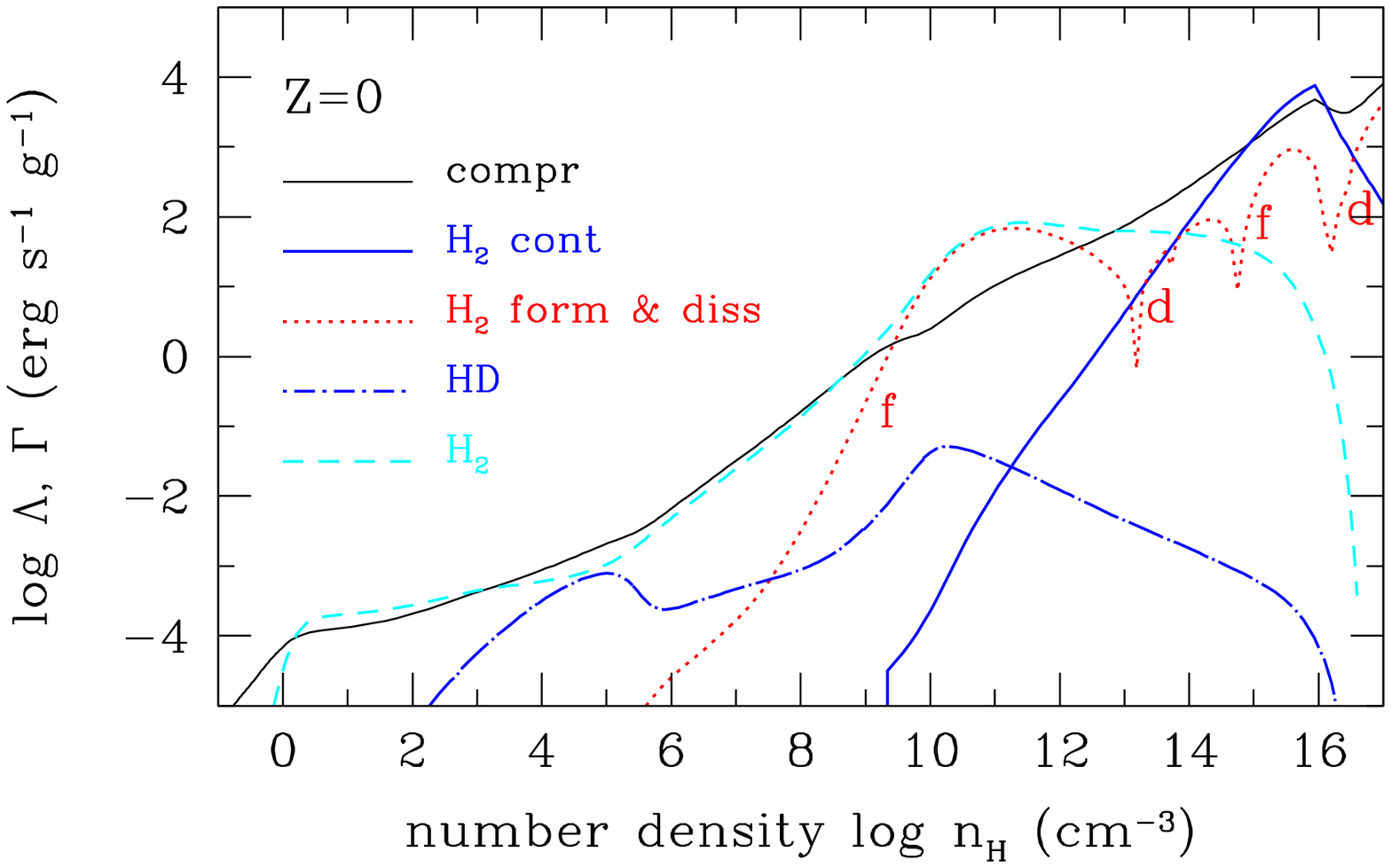}{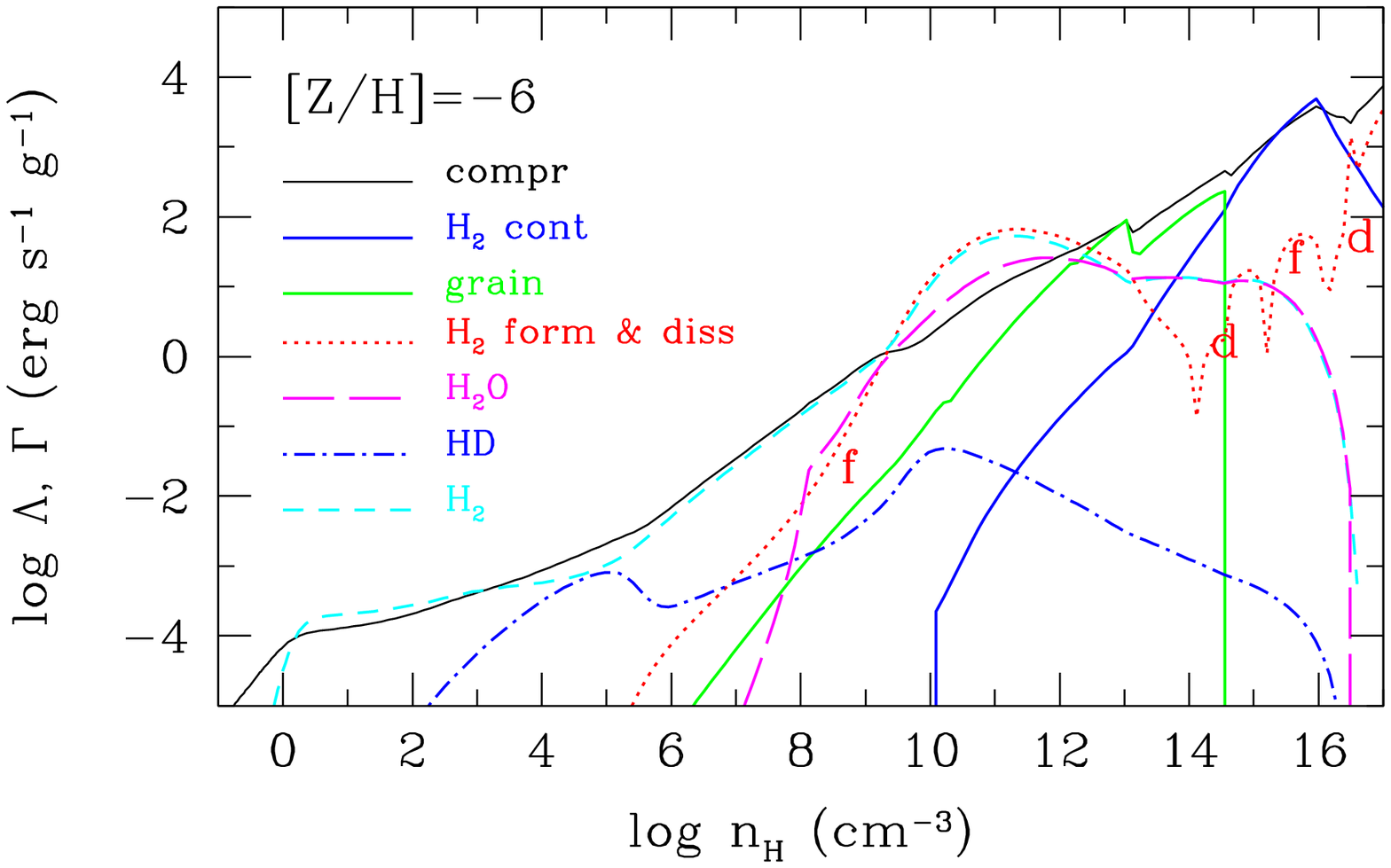}
\plottwo{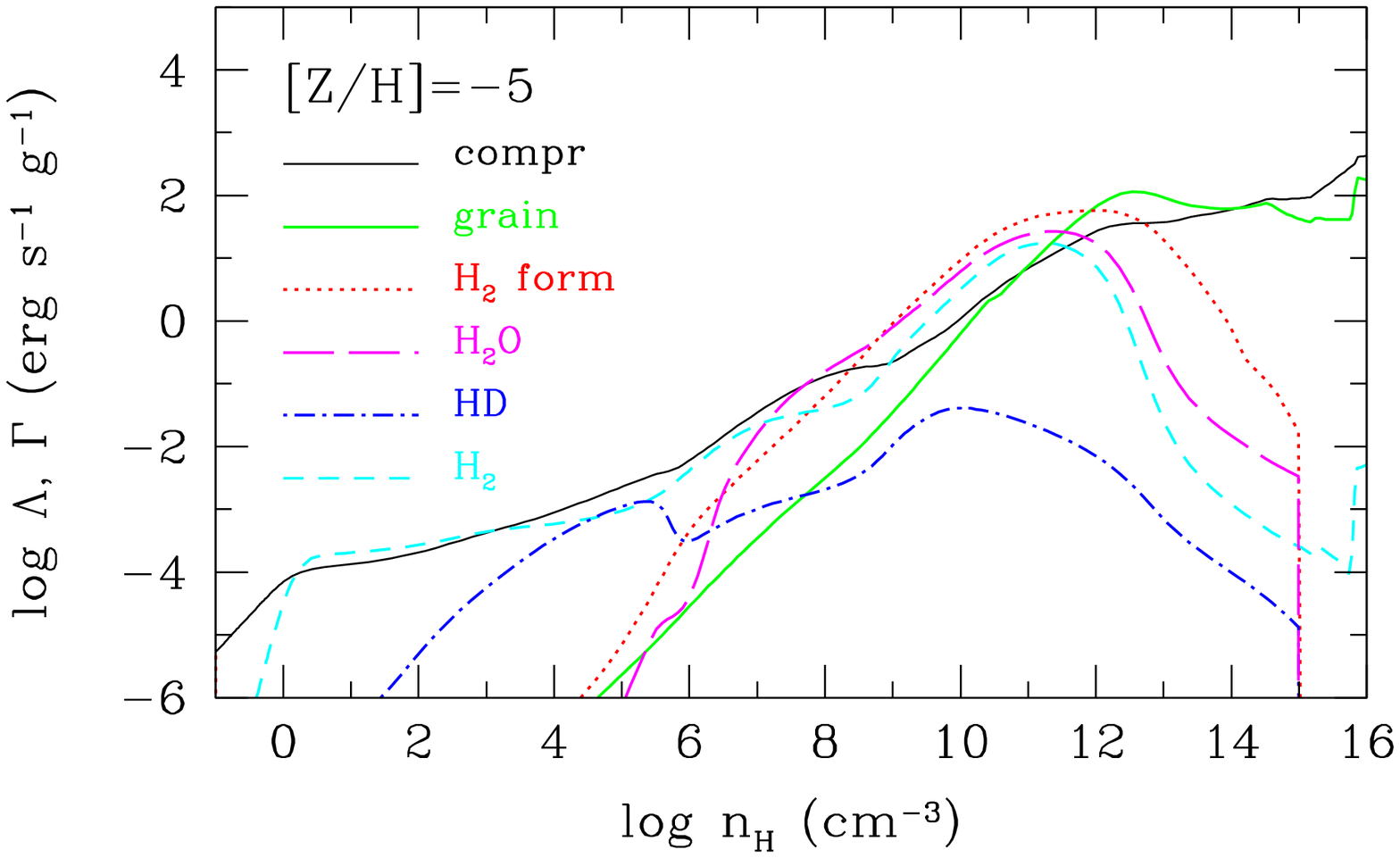}{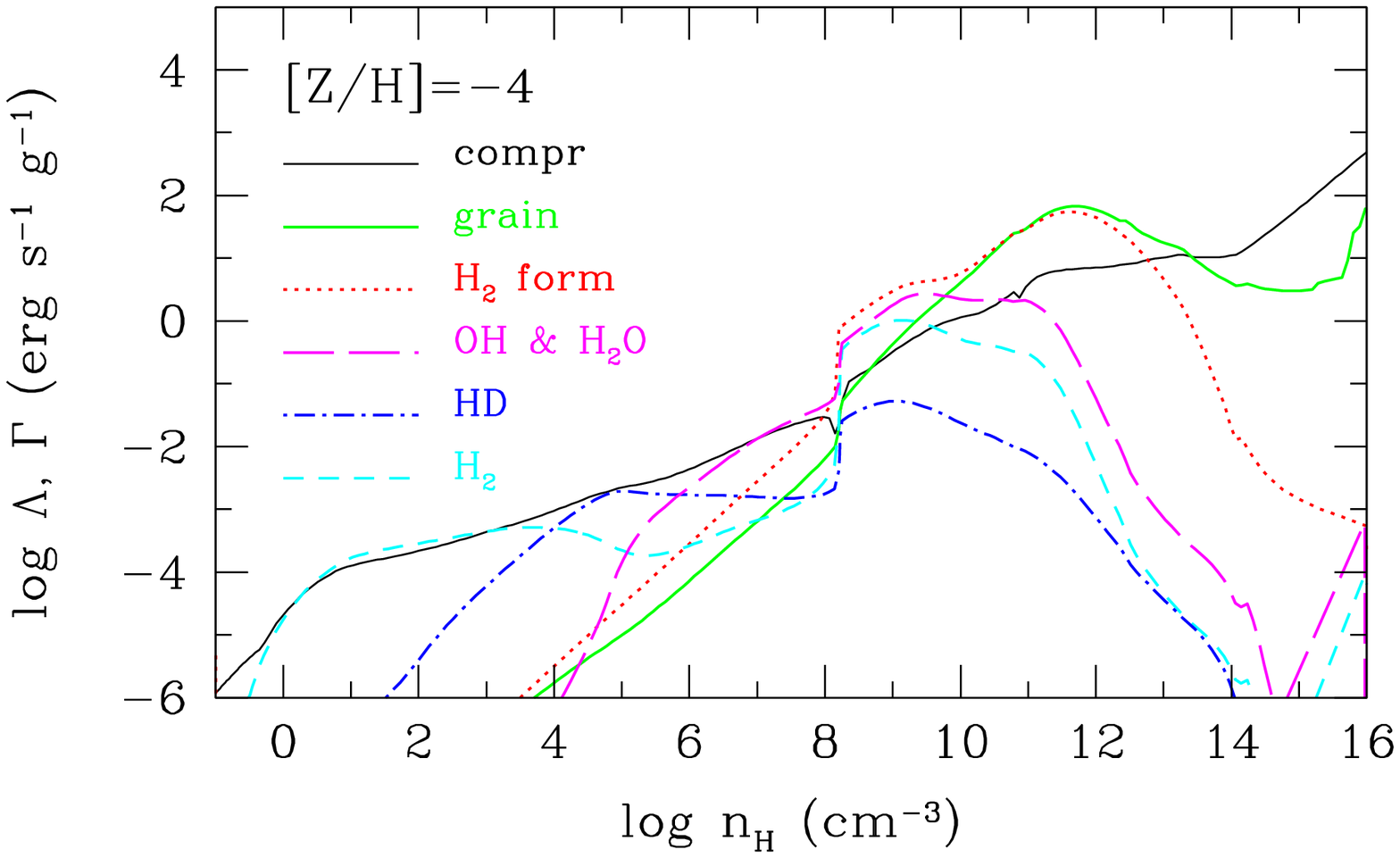}
\plottwo{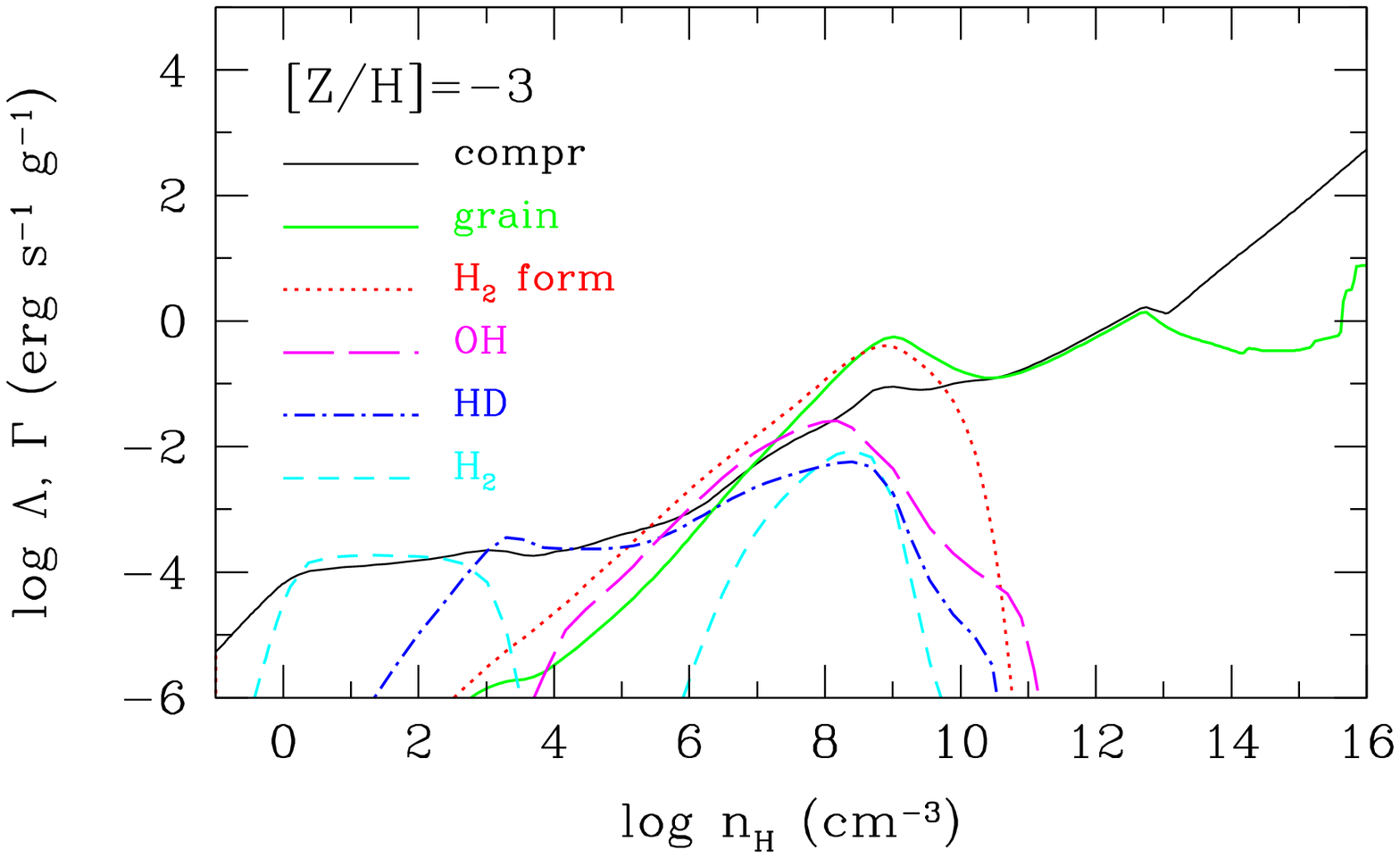}{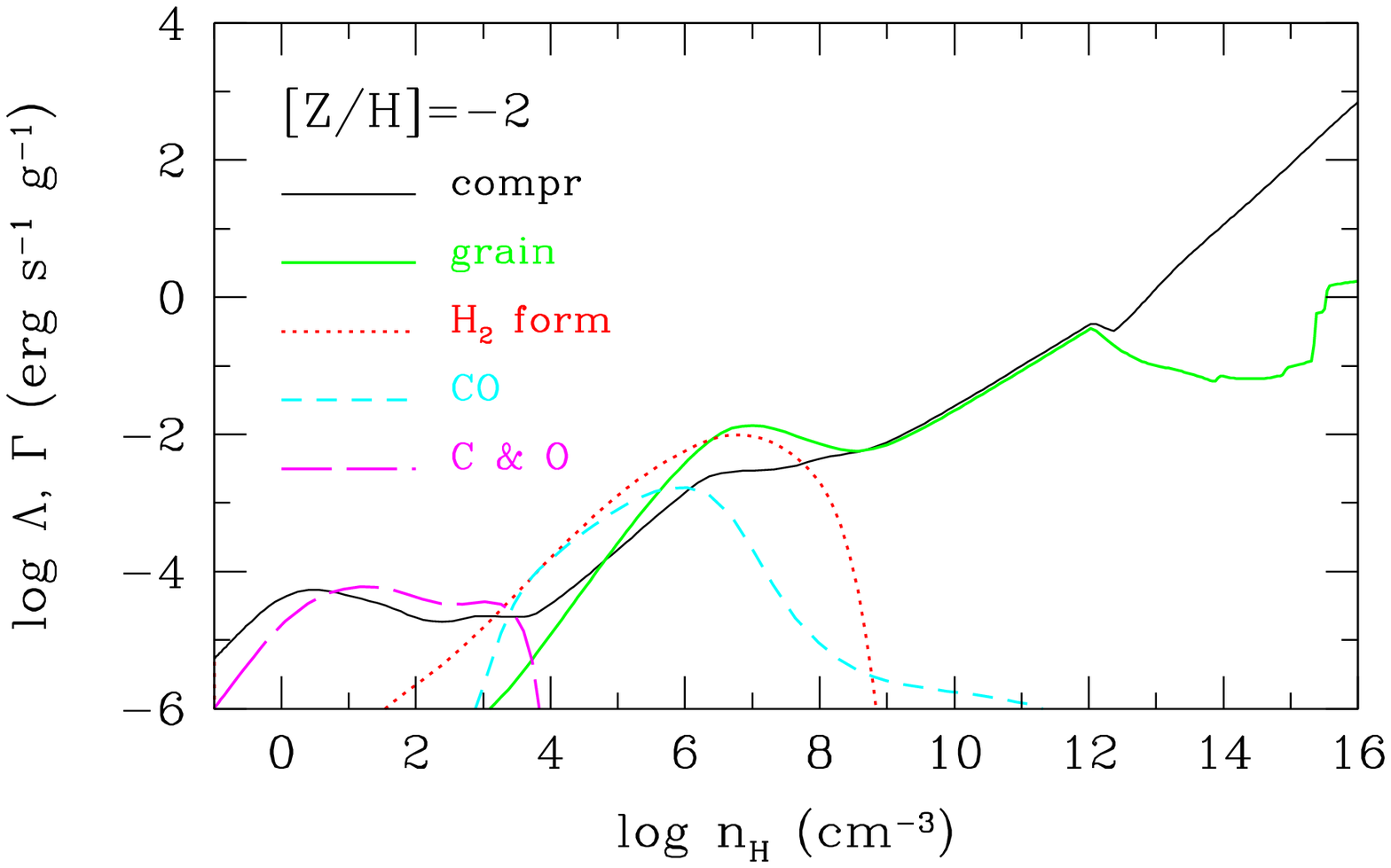}
\plottwo{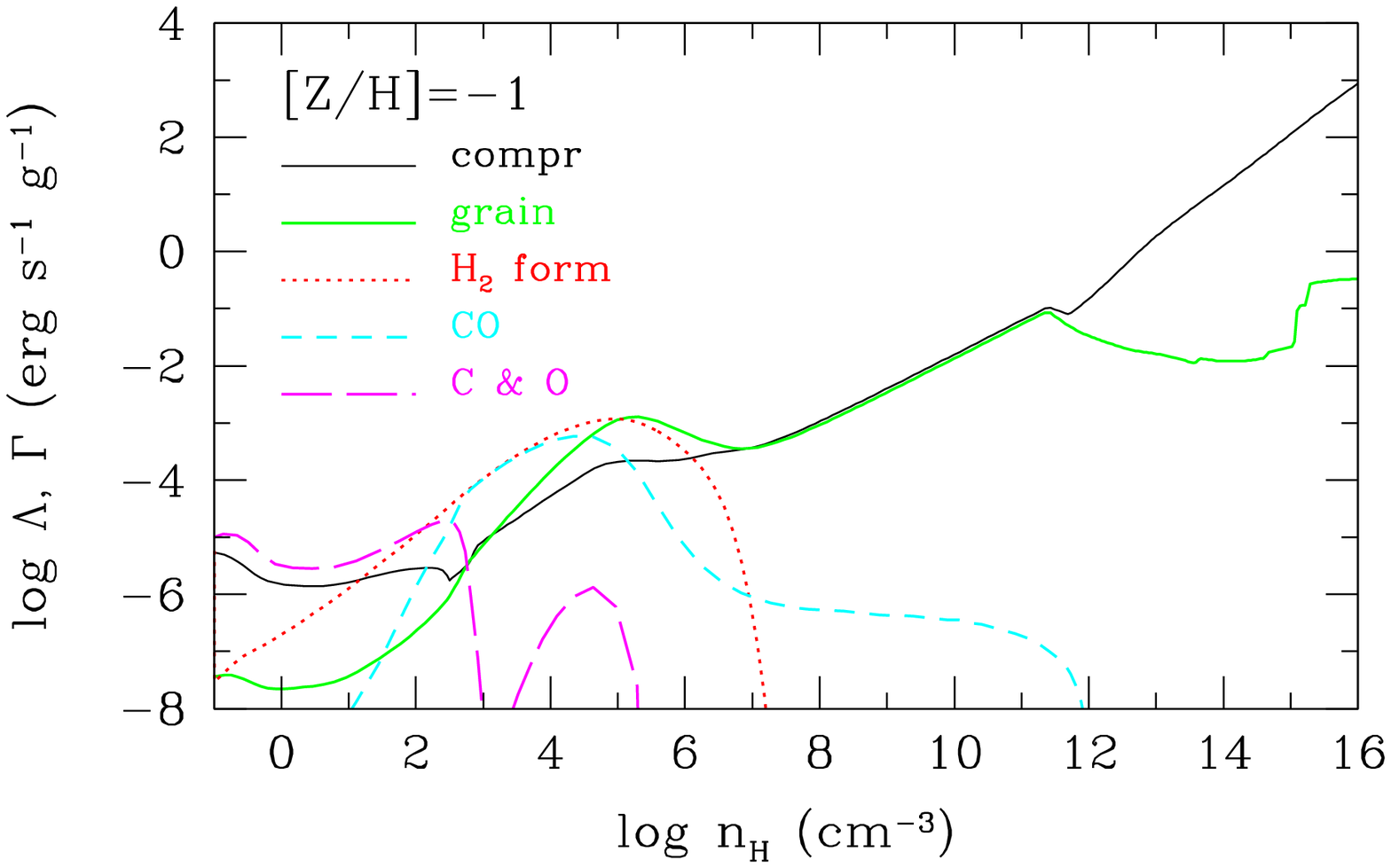}{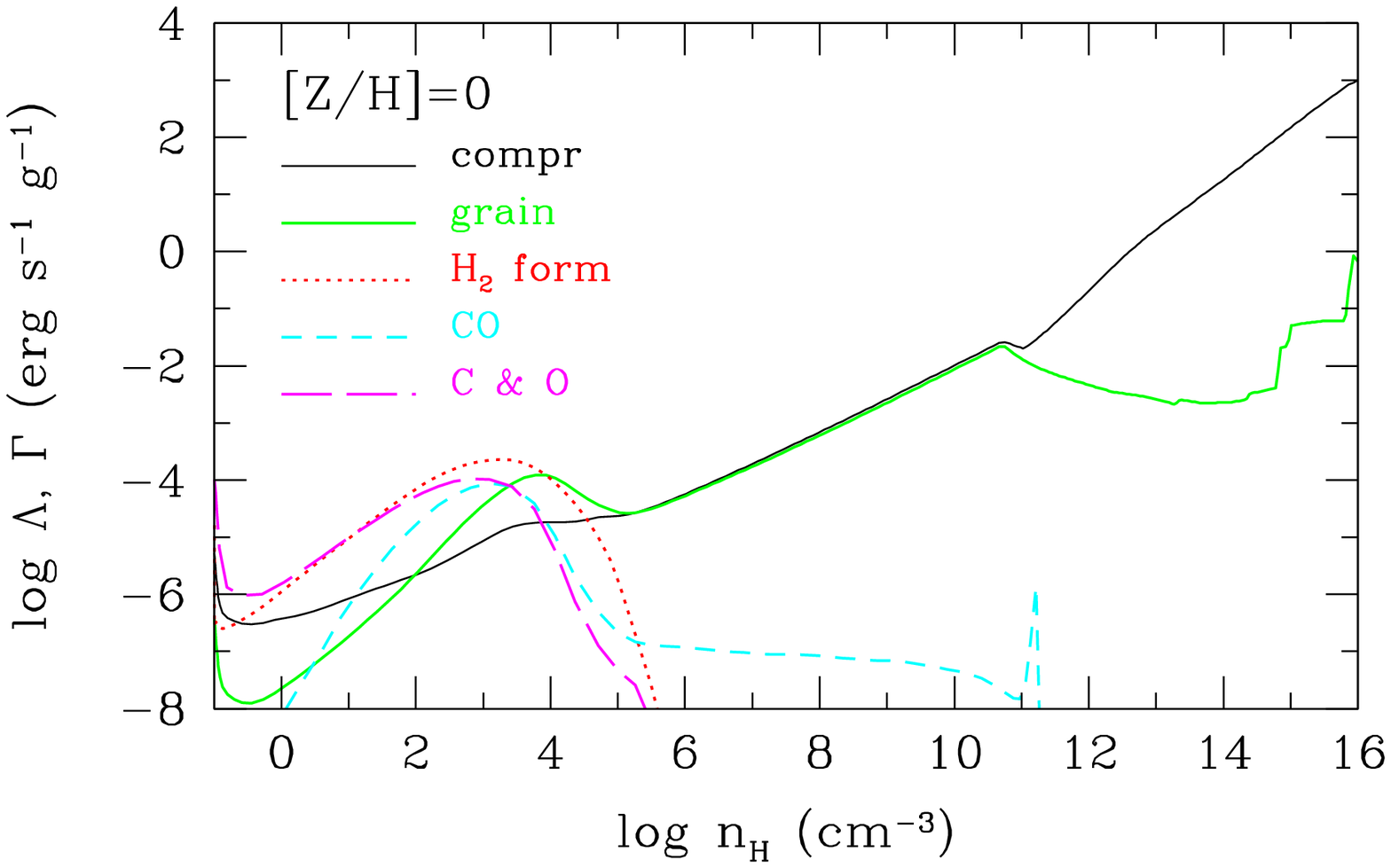}
\figcaption[cool]{Energy balance diagrams. 
Contribution to total cooling and heating rates (per unit mass) 
from each processes are shown as a function of number density $n_{\rm H}$ 
for fiducial cases with (a)[Z/H]=-$\infty$, (b) -6, (c) -5, (d) -4, (e) -3, 
(f) -2, (g) -1, and (h) 0.
The meaning of symbols are as follows:
``compr''; compressional heating, ``H$_2$ cont''; cooling by 
H$_2$ collision-induced emission, 
``grain''; cooling by dust thermal emission, 
``H$_2$ form (diss)''; H$_2$ formation heating (dissociation cooling, 
respectively), which is indicated by ``f''(``d'', respectively) in the Figure.
Others are line-emission-cooling rates by indicated coolants. 
In particular, ``[C\&O]'' means cooling by fine-structure lines 
of C$^{+}$, C, and O.
In panel (d), ``OH\&H$_2$O'' is cooling by OH for 
$n_{\rm H}\la 10^{8}{\rm cm^{-3}}$
and by H$_2$O for higher densities.
Note that the compression and the H$_2$ formation are heating mechanism, 
while the other processes are cooling mechanism.
\label{fig:cool}}

\plotone{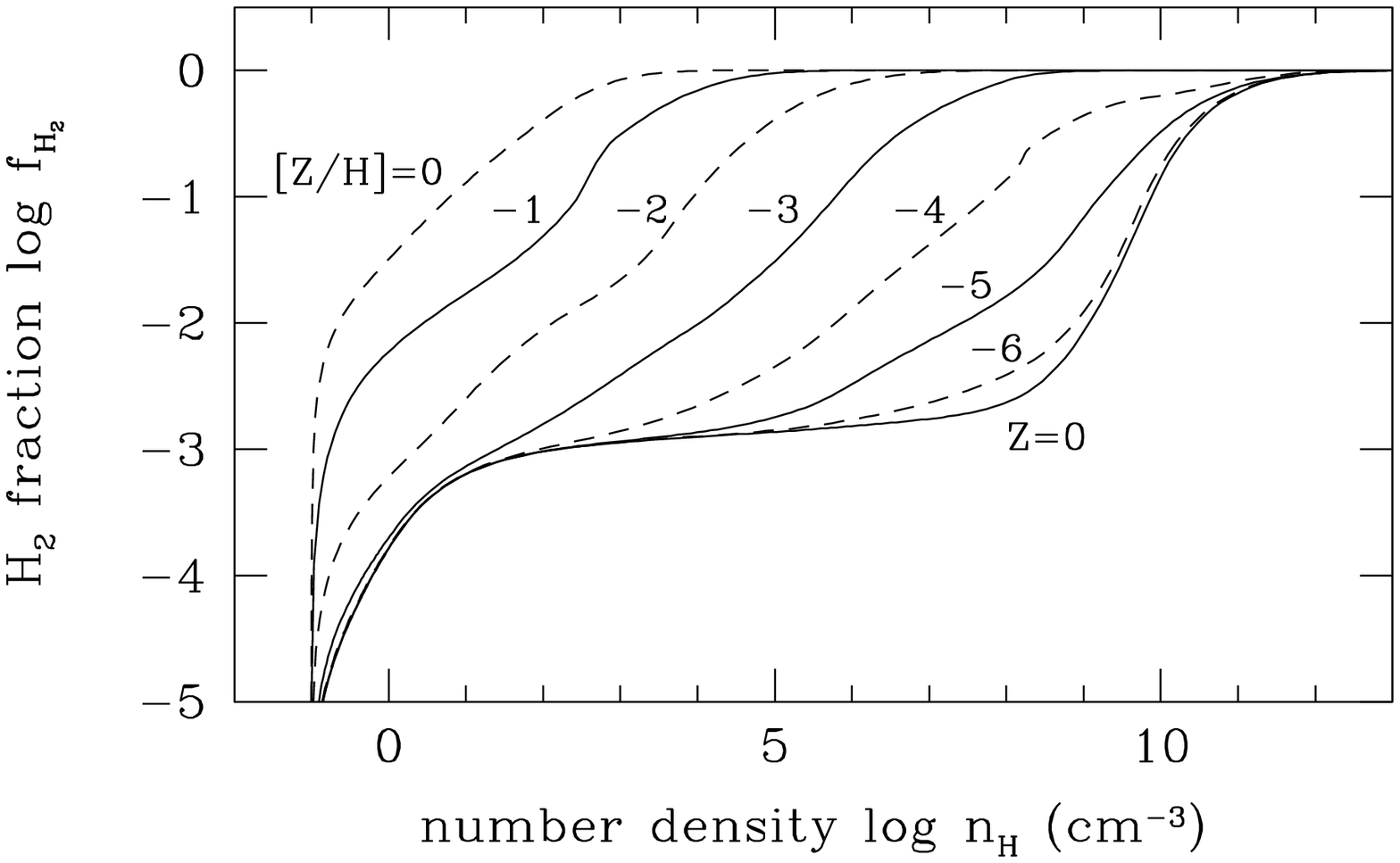}
\figcaption[fH2]{Evolution of H$_2$ fraction $f_{\rm H_2} \equiv 
2 y({\rm H_2})$ for the cases shown in Figure 1.
Solid (dashed) lines indicate those for [Z/H]=-$\infty$ (Z=0), -5, -3, and -1
([Z/H]=-6, -4, -2, and 0, respectively).
Note that $f_{\rm H_2}=1$ means that hydrogen is fully molecular.
\label{fig:fH2}}

\plotone{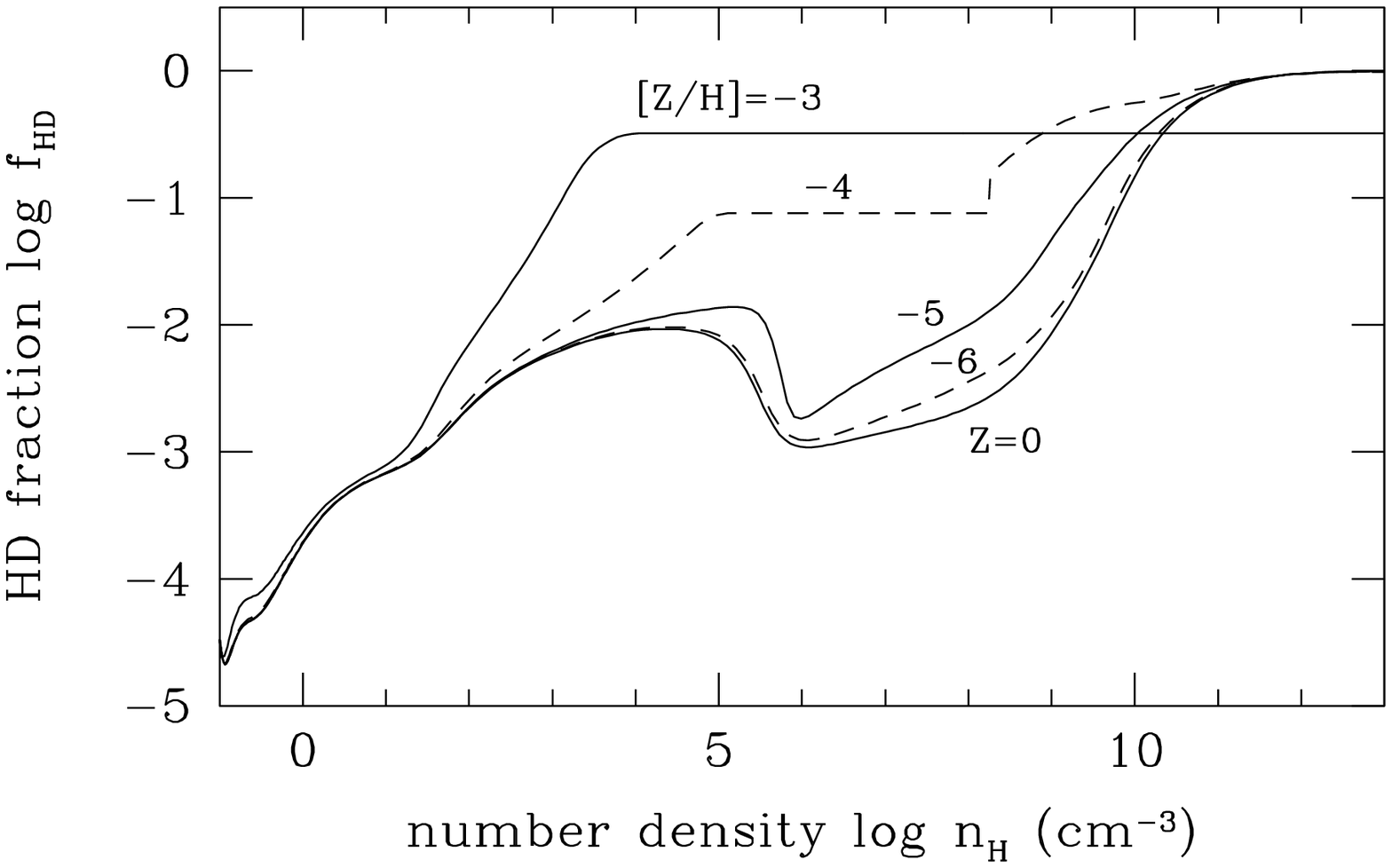}
\figcaption[fHD]{Evolution of HD fraction 
$f_{\rm HD} \equiv y({\rm HD})/y_{\rm D}$ in low-metallicity clouds. 
Solid (dashed) lines show the cases of [Z/H]=- $\infty$ (Z=0), -5, and -3
([Z/H]=-6, and -4, respectively). 
Note that $f_{\rm D}=1$ means that deuterium is fully molecular.
\label{fig:fHD}}

\plottwo{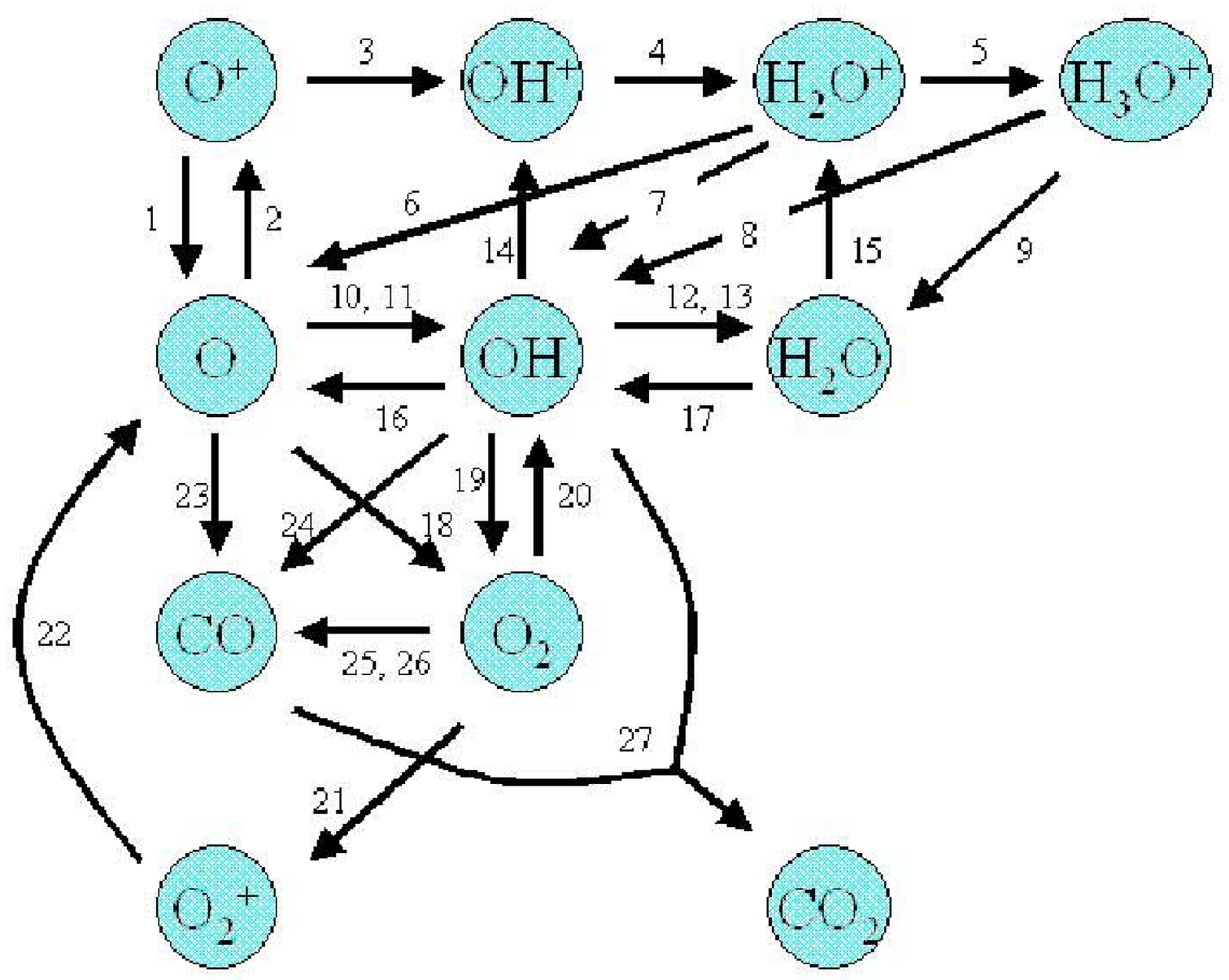}{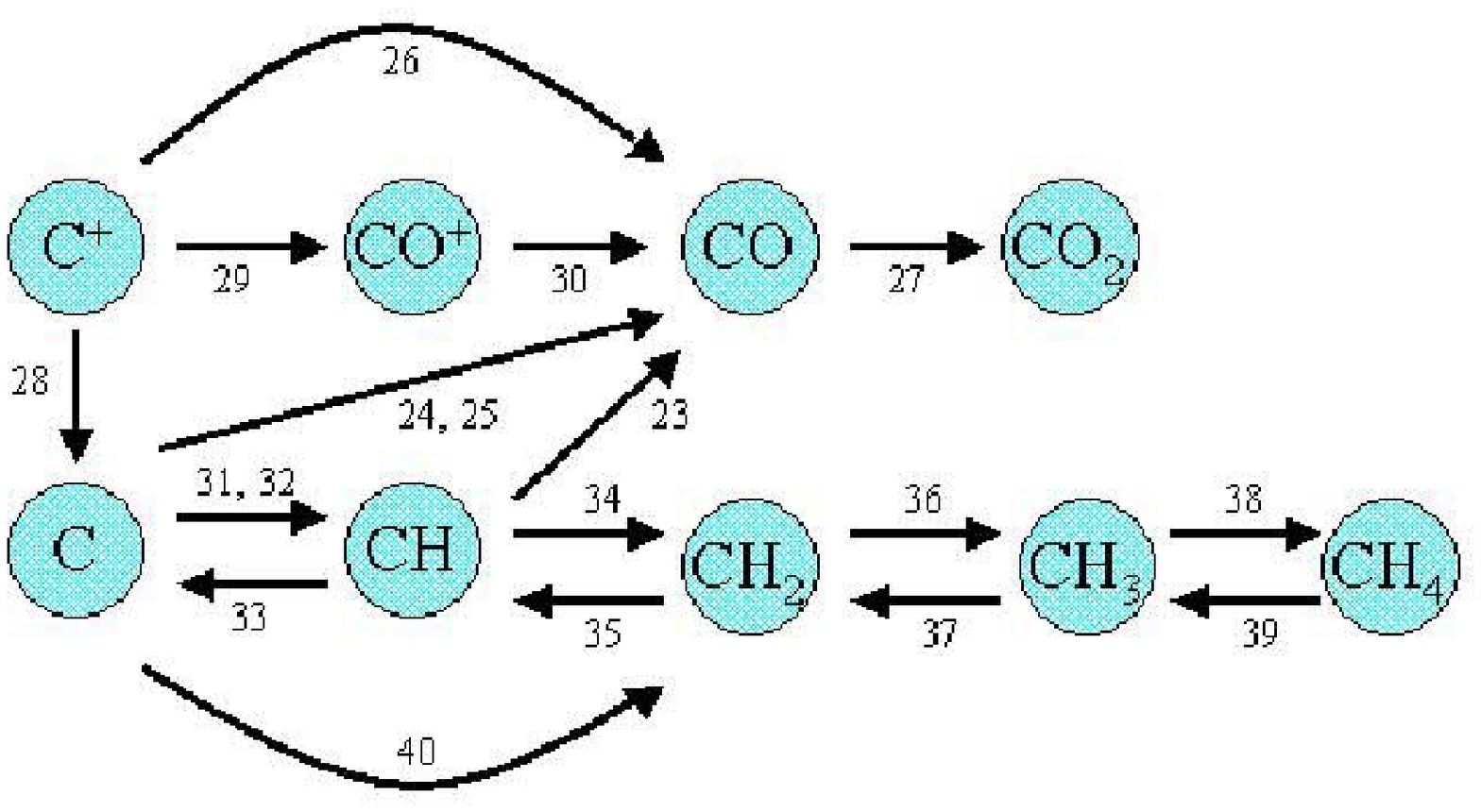}
\figcaption[COchem]{Important chemical network of (a) oxygen (b) carbon. 
All the C and O related reactions in the reduced network (see text) are 
shown. 
The indicated numbers are the same as Z reactions in Table 1, 
i.e., $i$ is for reaction Z$i$. 
\label{fig:COchem}}

\plottwo{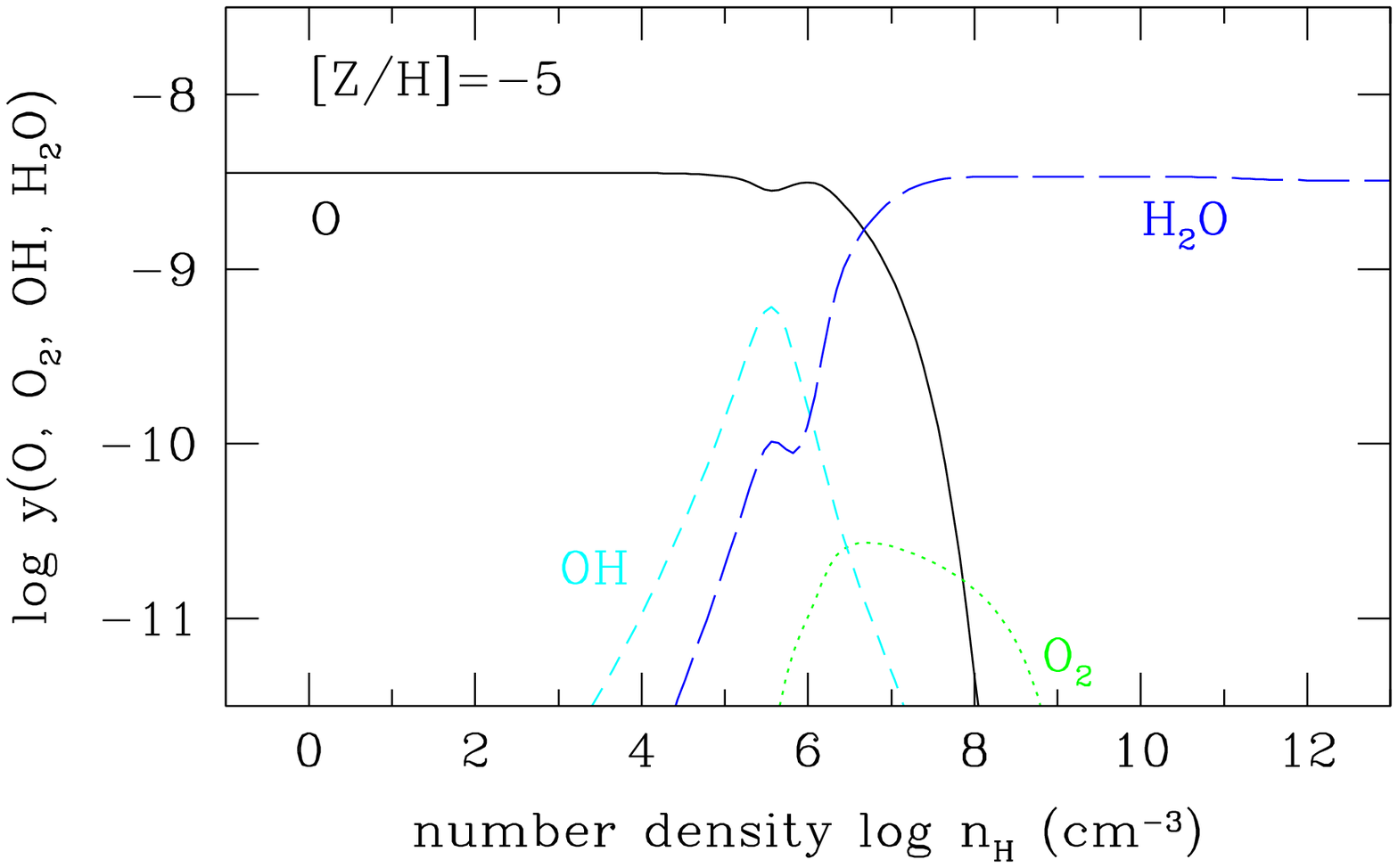}{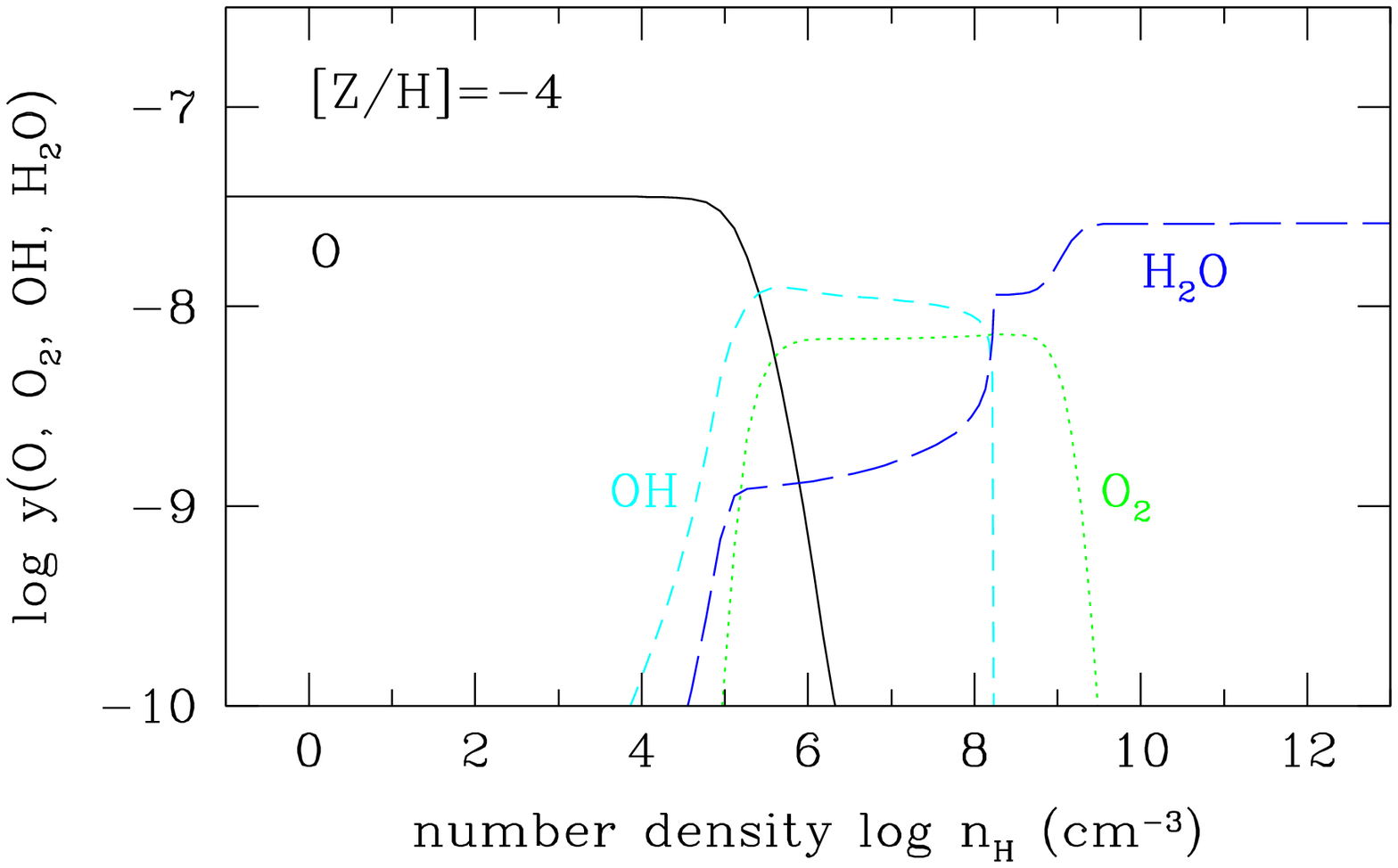}
\plottwo{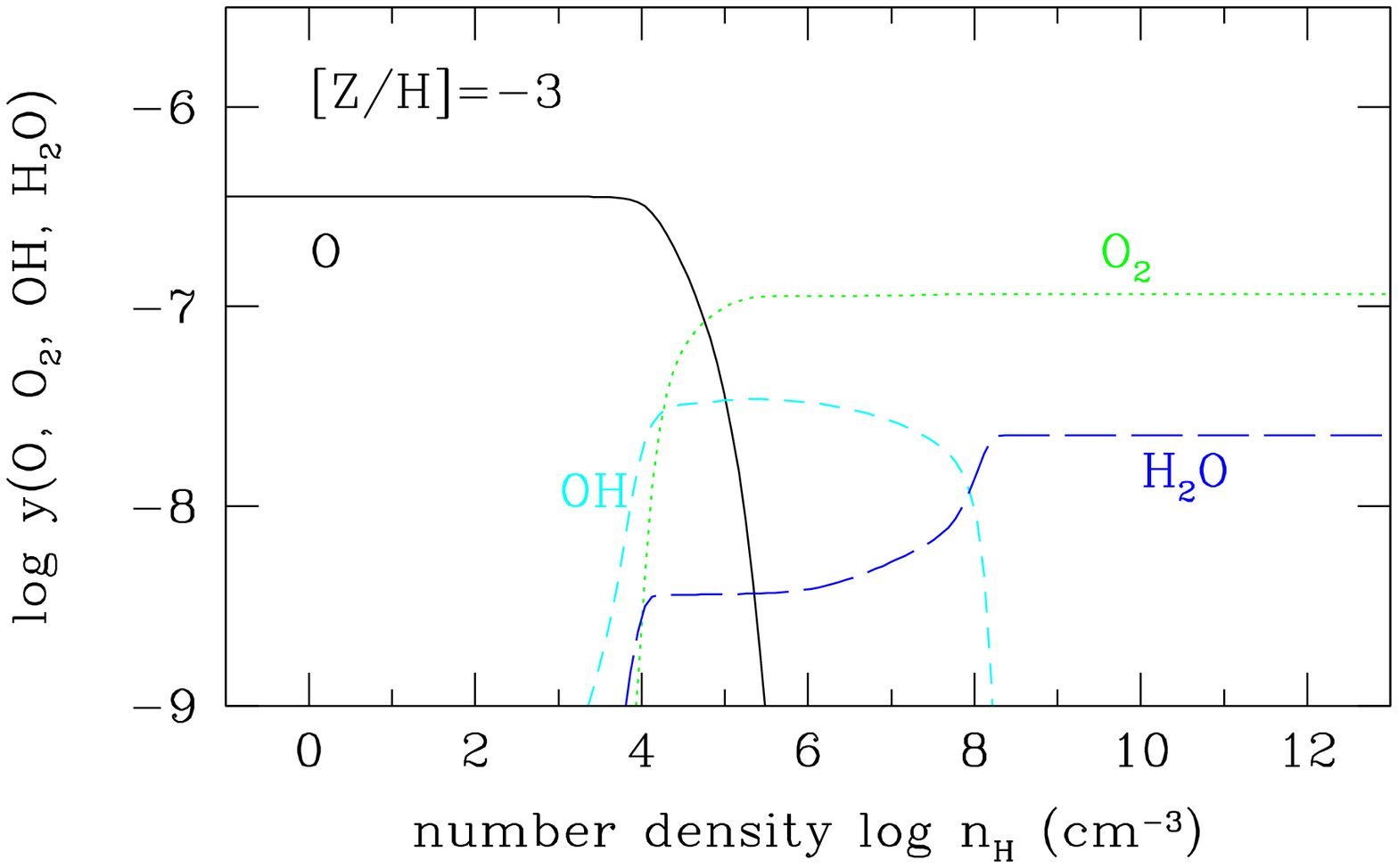}{white.eps}
\figcaption[yO]{Transformation of oxygen in metal-deficient clouds with
metallicity (a)[Z/H]=-5, (b) -4, and (c) -3 in the fiducial cases.
Concentration of species O, OH, H$_2$O, and O$_2$ are shown against H number
density $n_{\rm H}$.
\label{fig:yO}}

\plotone{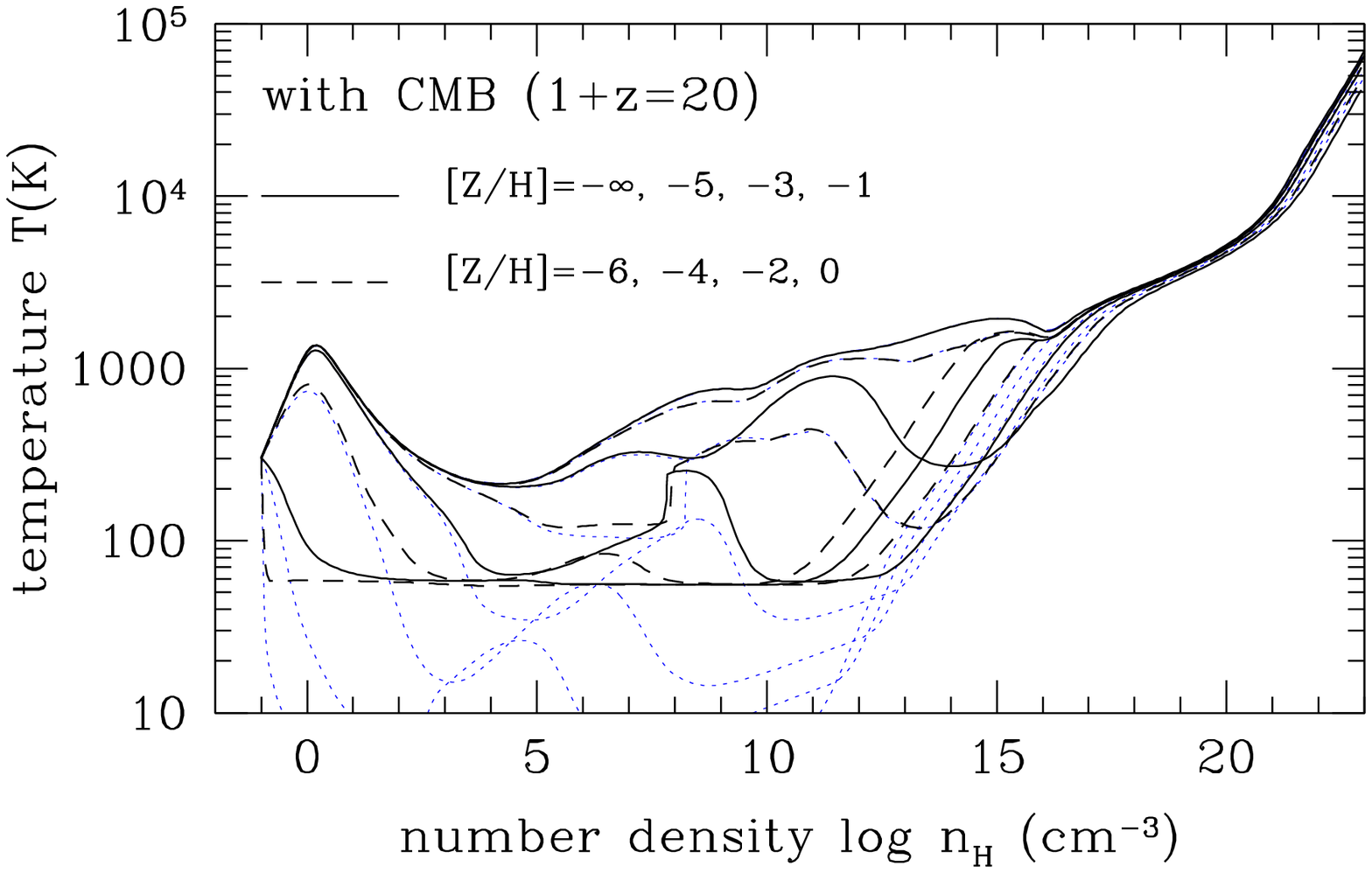}
\figcaption[nT.Tr60]{Same as Figure 1, but with external radiation of 
CMB at 1+z=20, i.e., radiation temperature $T_{\rm rad}=54.6$K.
Those of the fiducial cases are shown by dotted lines for comparison.
\label{fig:nT.Tr60}}

\plotone{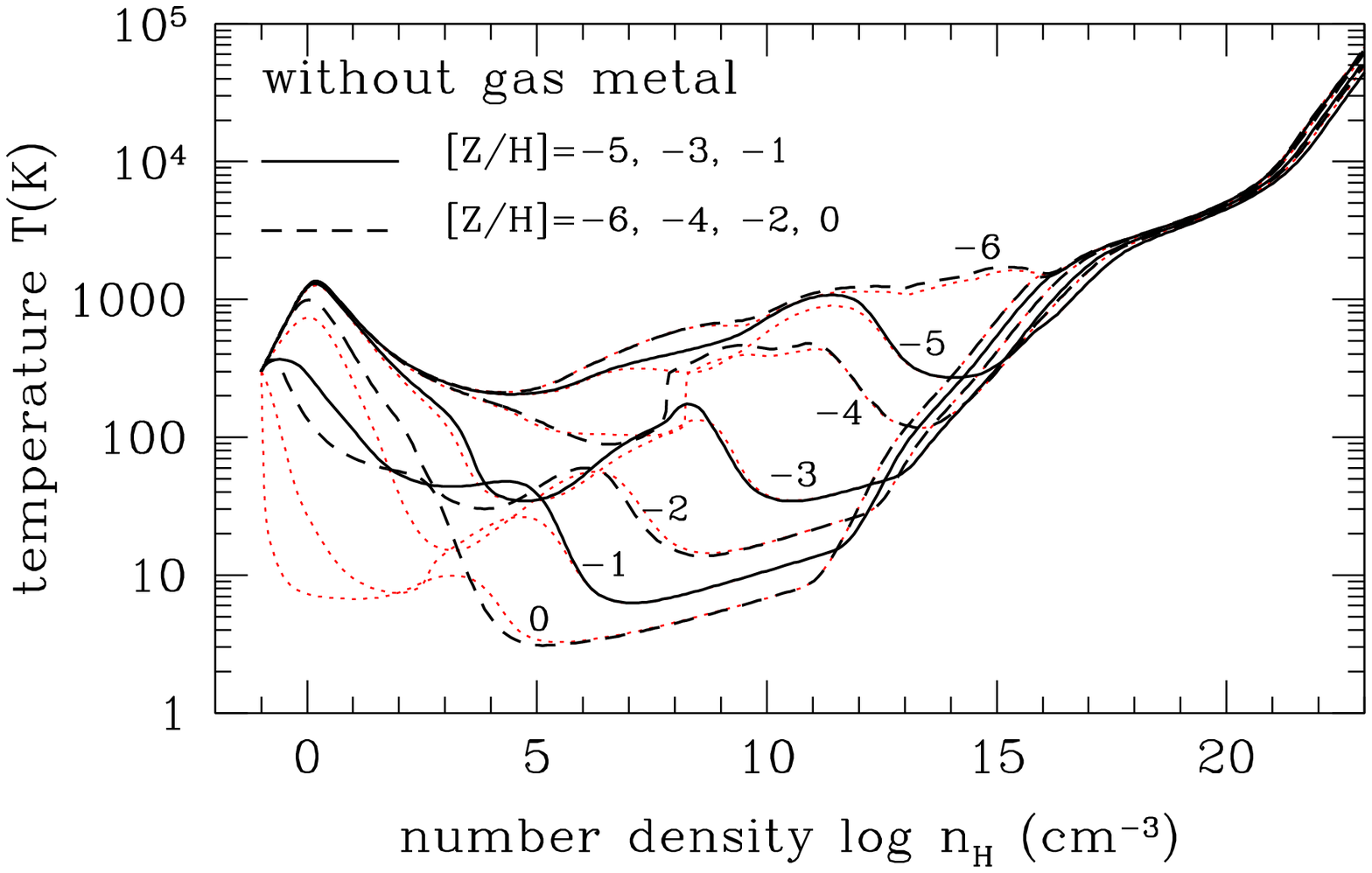}
\figcaption[nT.dustonly]{Same as Figure 1, but without gas-phase metals.
The amount of dust is equal to the fiducial case with the same value of 
[Z/H] indicated.
Also shown by dotted lines are those of the fiducial cases.
Note that the total amount of metals are smaller than in the fiducial case 
with the same [Z/H] value because of the lack of gas-phase metals.
\label{fig:nT.dustonly}}

\plotone{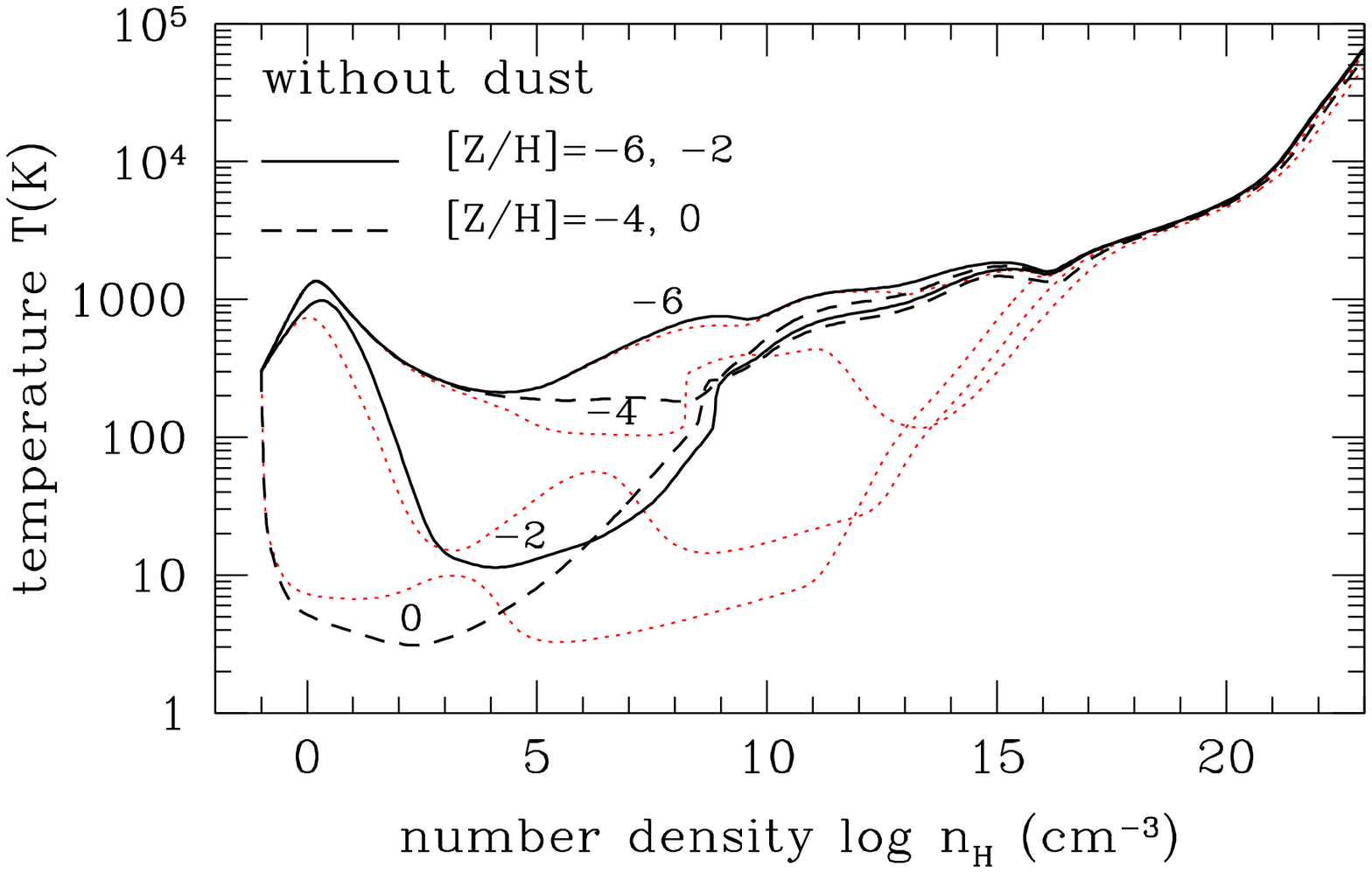}
\figcaption[nT.nodust]{Same as Figure 1, but without dust grains.
For clarity, fewer cases are shown here than in Figure 1.
An amount of metals in gas phase is the same as the fiducial cases.
Also shown by dotted lines are those of the fiducial cases.
Note that the total amount of metals are smaller than in the fiducial case 
with the same [Z/H] value because of the lack of metals in dust.
\label{fig:nT.nodust}}

\plotone{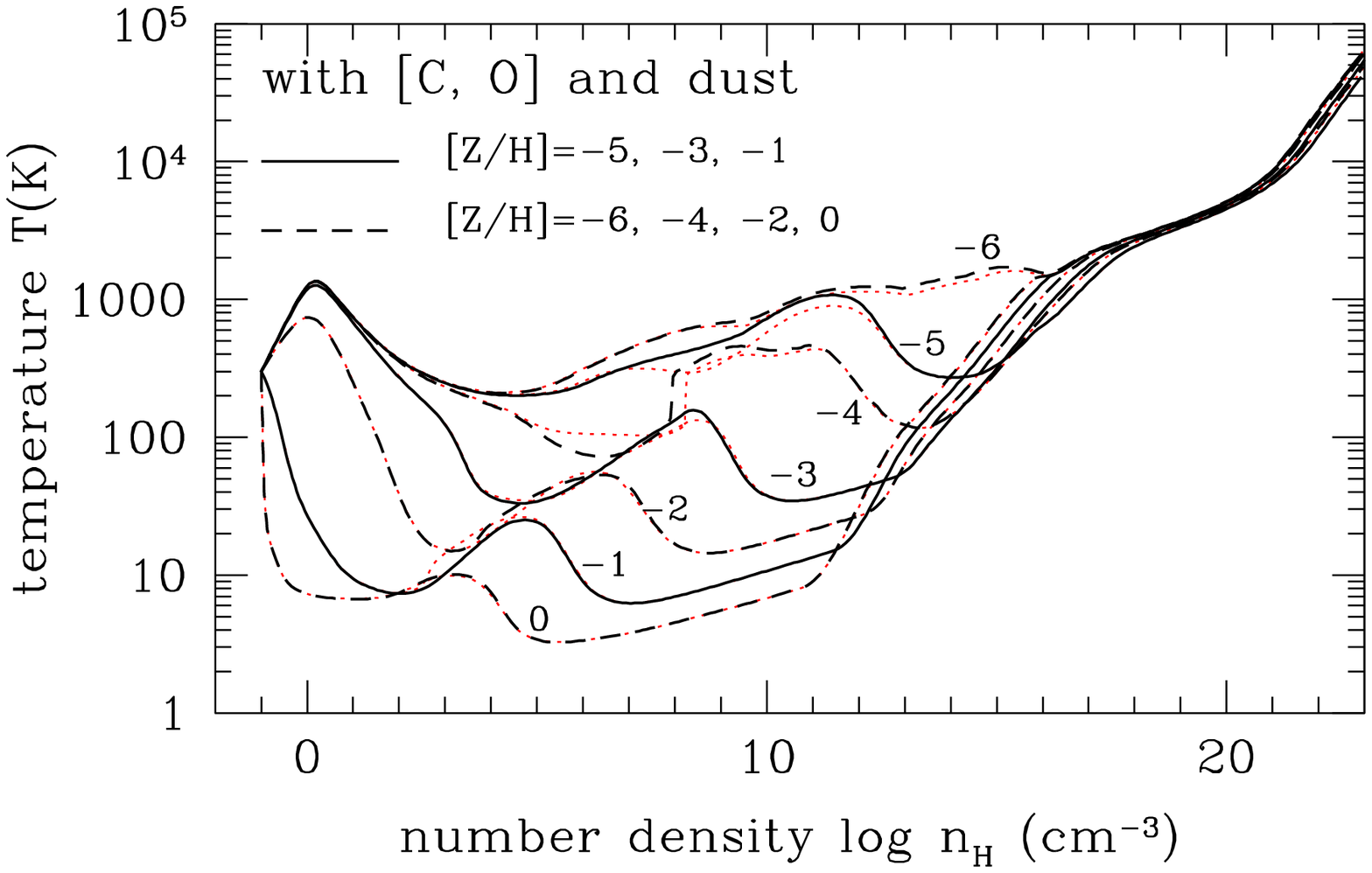}
\figcaption[nT.smchem]{Temperature evolution calculated by the minimum model, 
which includes fine-structure-line cooling and dust processes only, 
in addition to the metal-free gas microphysics.
Among metal reactions, only the recombination of C (reaction Z28 in Table 1) 
are included.
Also shown by dotted lines are those by the full network (the fiducial cases).
\label{fig:nT.smchem}}

\plotone{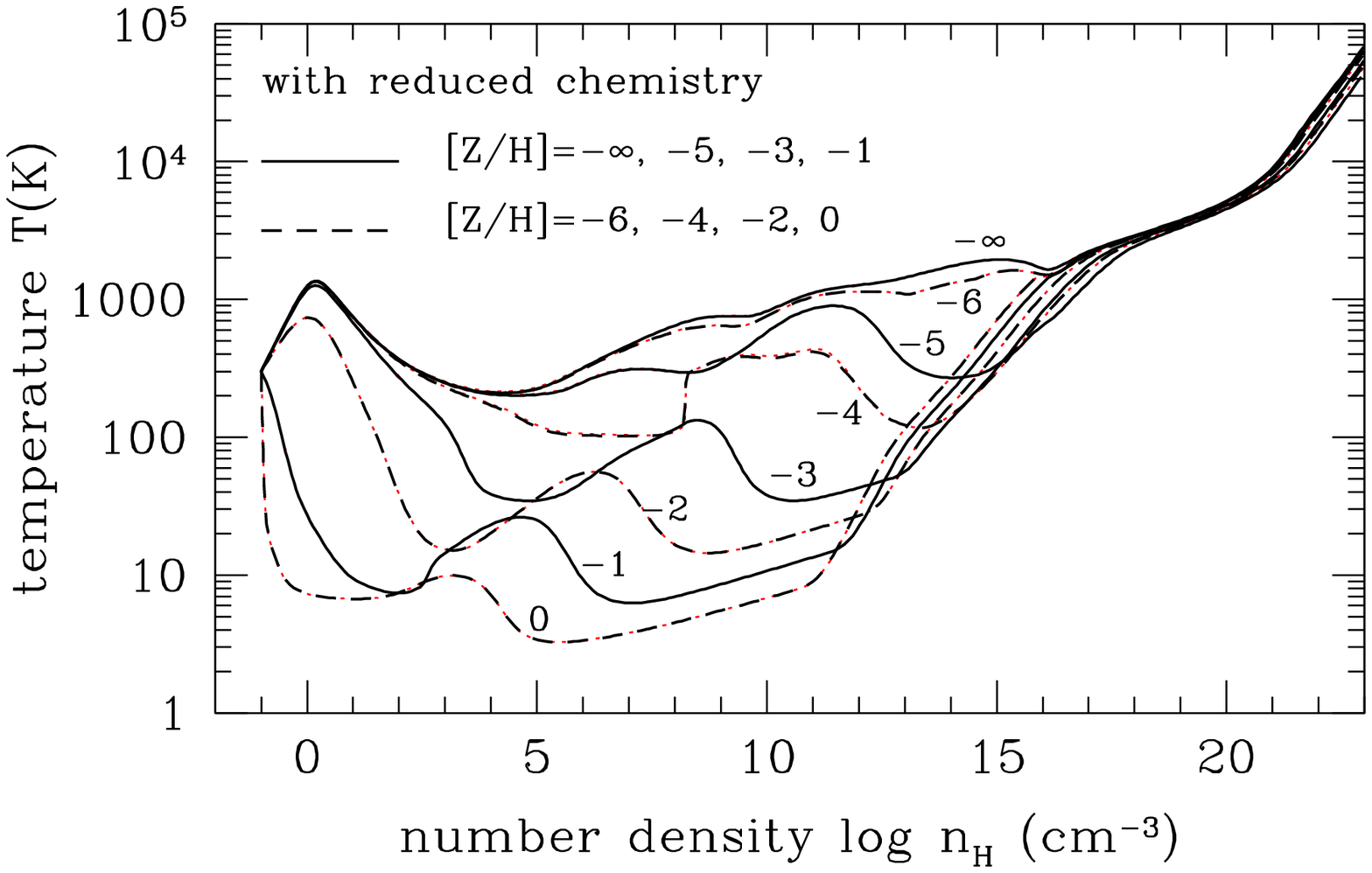}
\figcaption[nT.minichem]{Temperature evolution with the reduced chemical 
network. The chemical reactions shown in Table 1 are included.
Although those with full network are shown by dotted lines, 
all the lines are overlapping and cannot be recognized.
\label{fig:nT.minichem}}

\plottwo{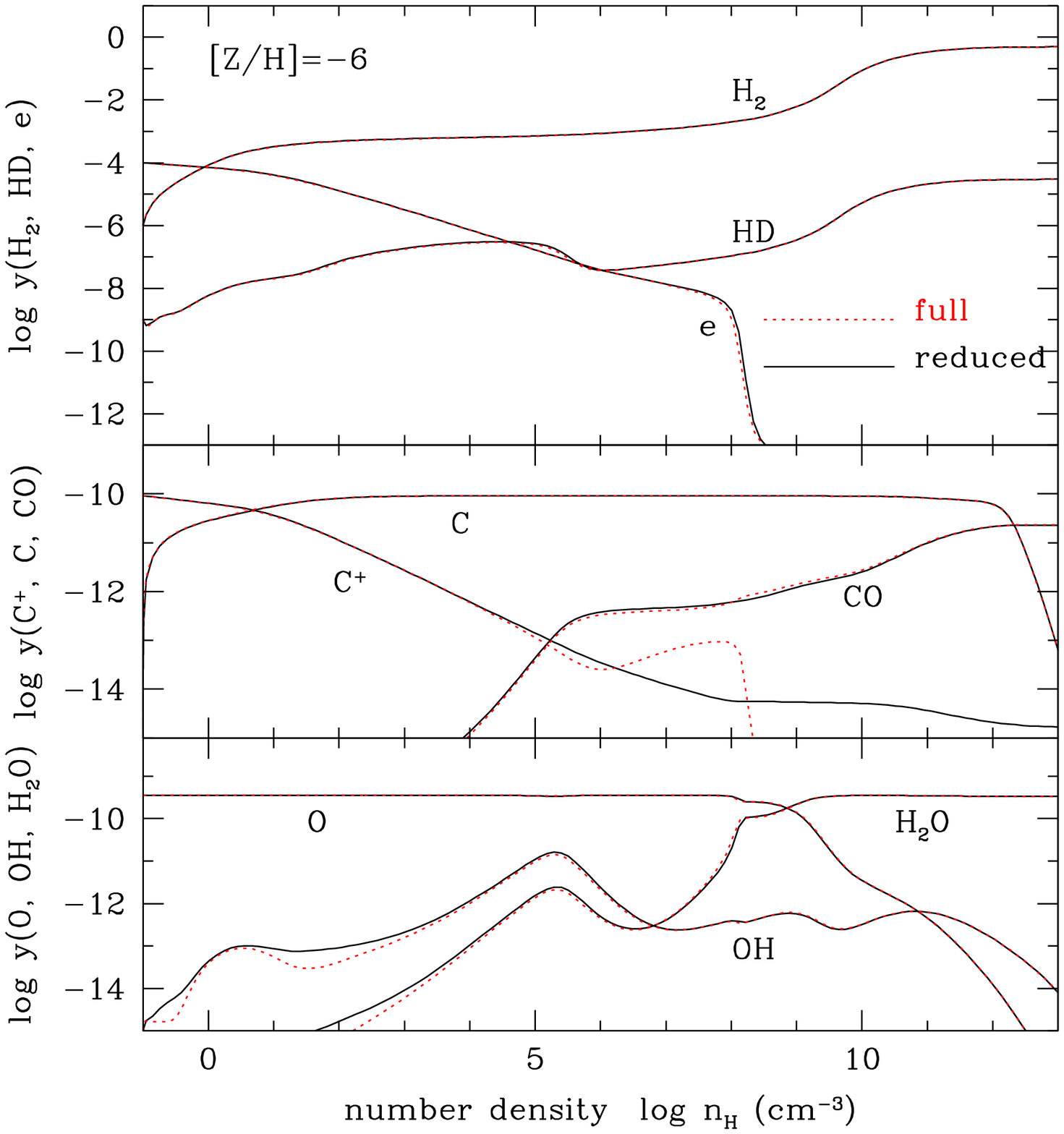}{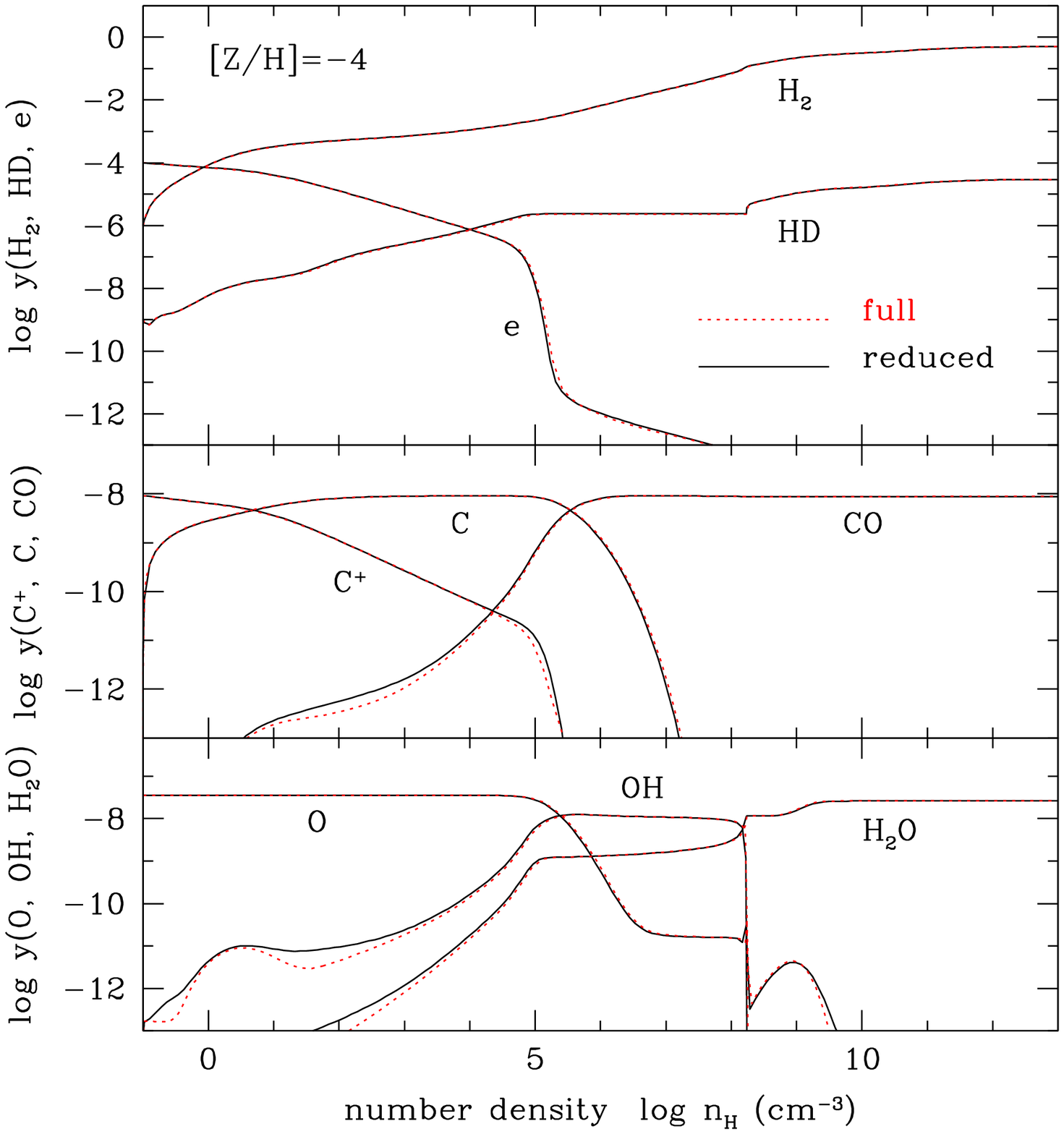}
\plottwo{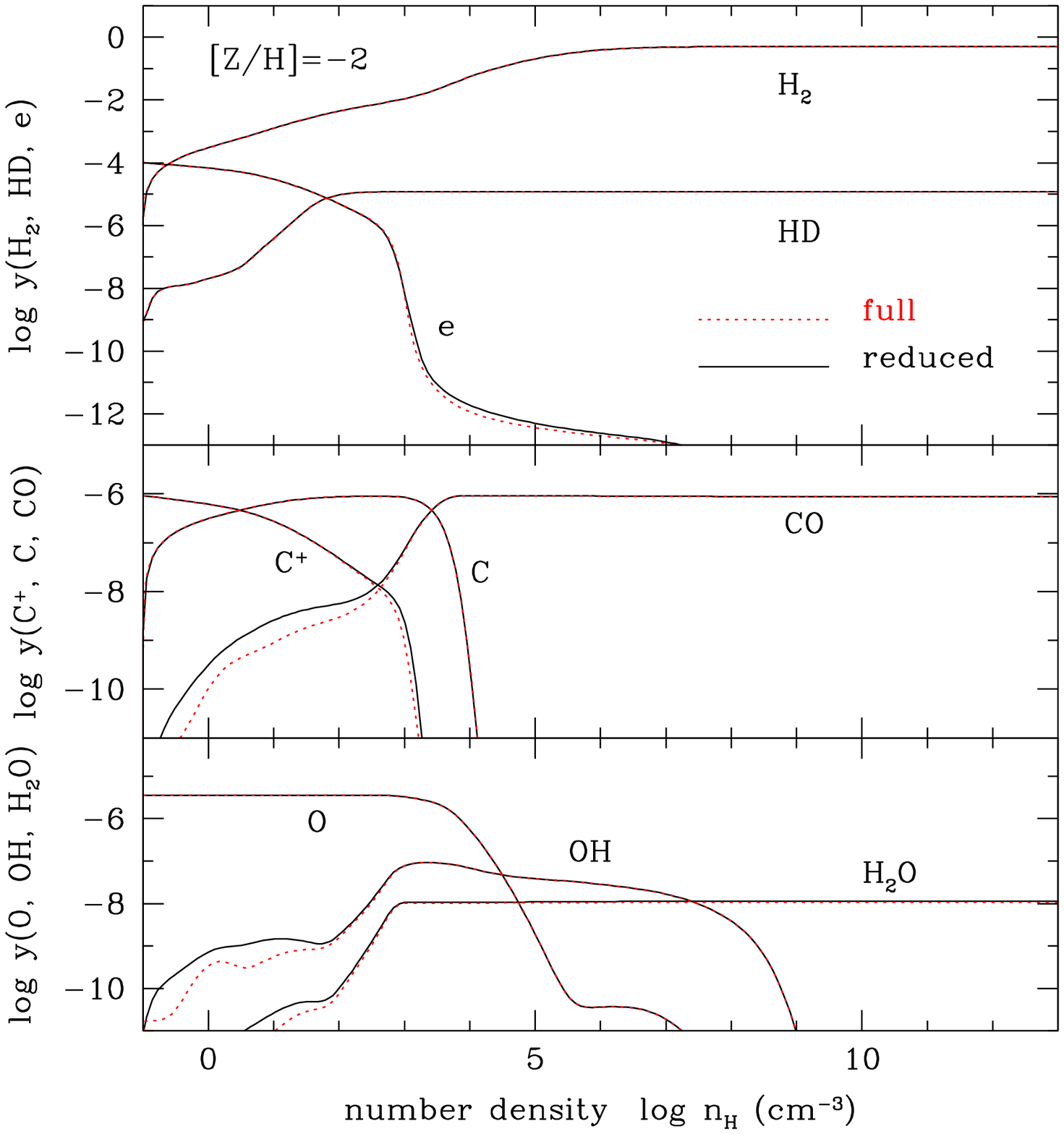}{white.eps}
\figcaption[ymini]{Abundances of important species by the reduced chemical 
network. Those with full network are shown by dotted lines.
\label{fig:ymini}}

\plottwo{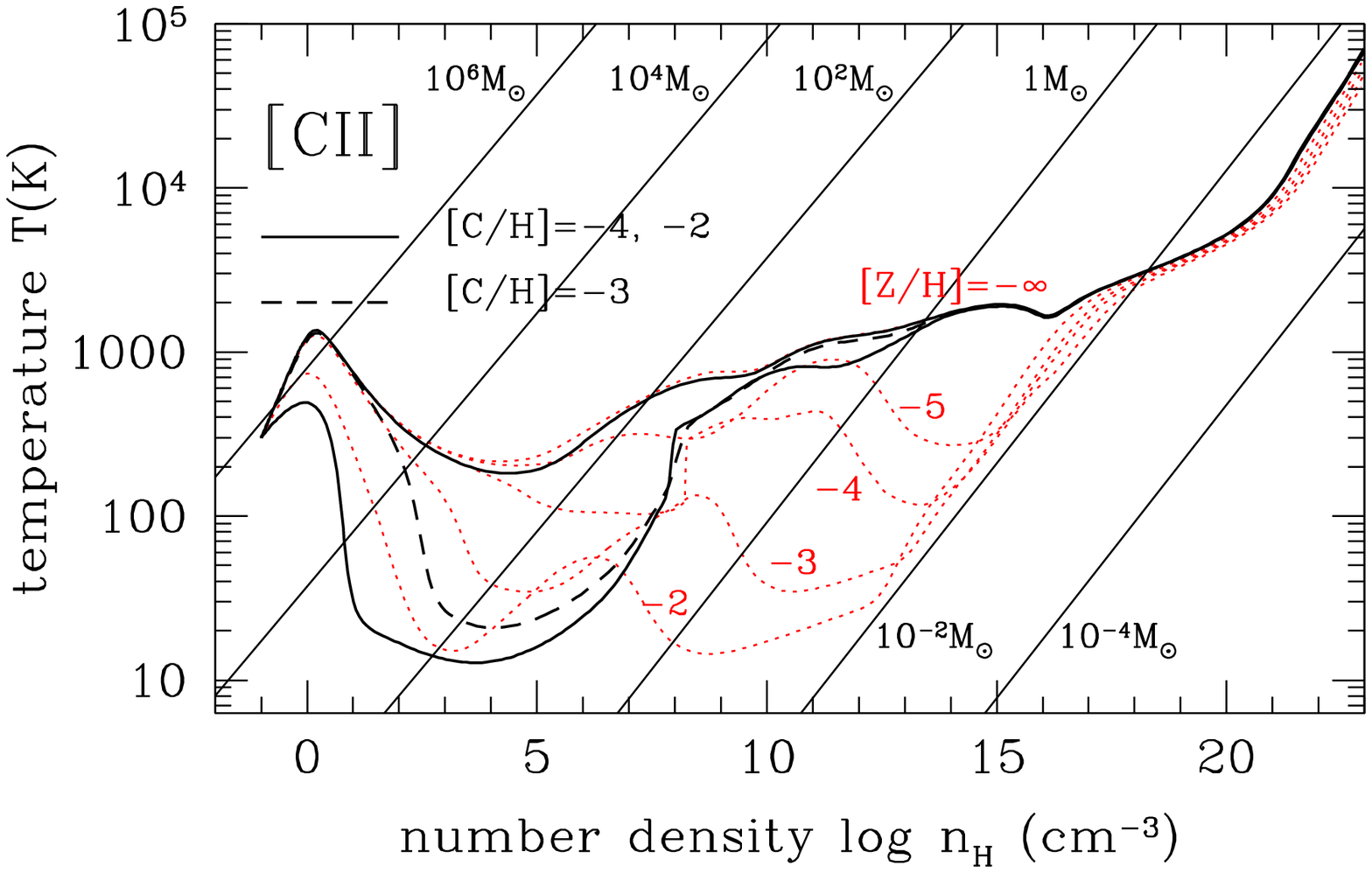}{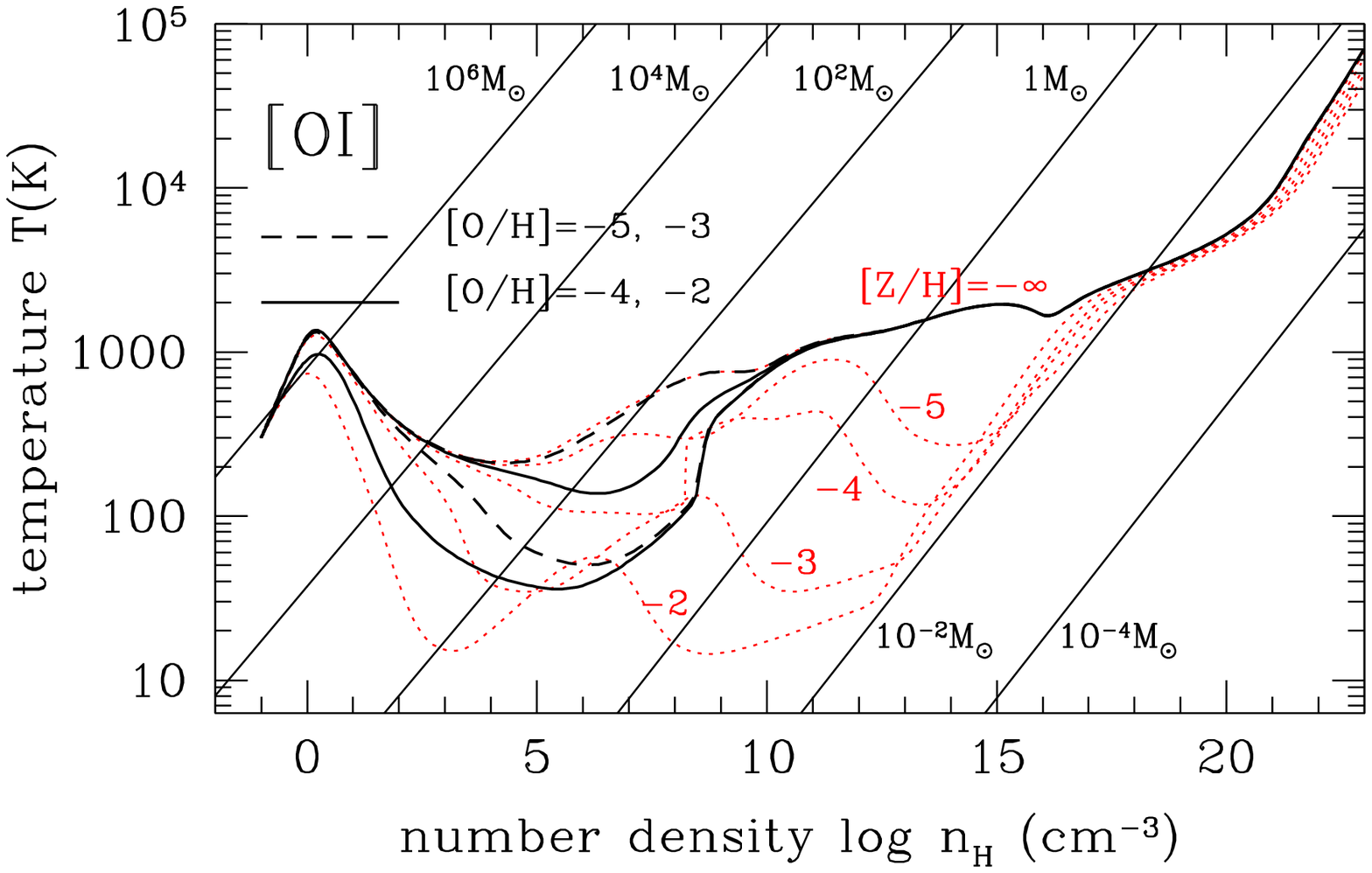}
\figcaption[nT.BL]{Temperature evolution calculated by the model 
that includes only (a)[CII] and (b)[OI] cooling in addition to 
the metal-free processes.
All C or O are assumed to be in the gas phase as in Bromm \& Loeb (2003).
The abundance of CII and OI are assumed to be constant in time:  
the metal chemistry is not solved.
Also shown by dotted lines are the fiducial cases with 
[Z/H]= $-\infty, 5, 4, 3, 2$.
\label{fig:nT.BL}}

\plotone{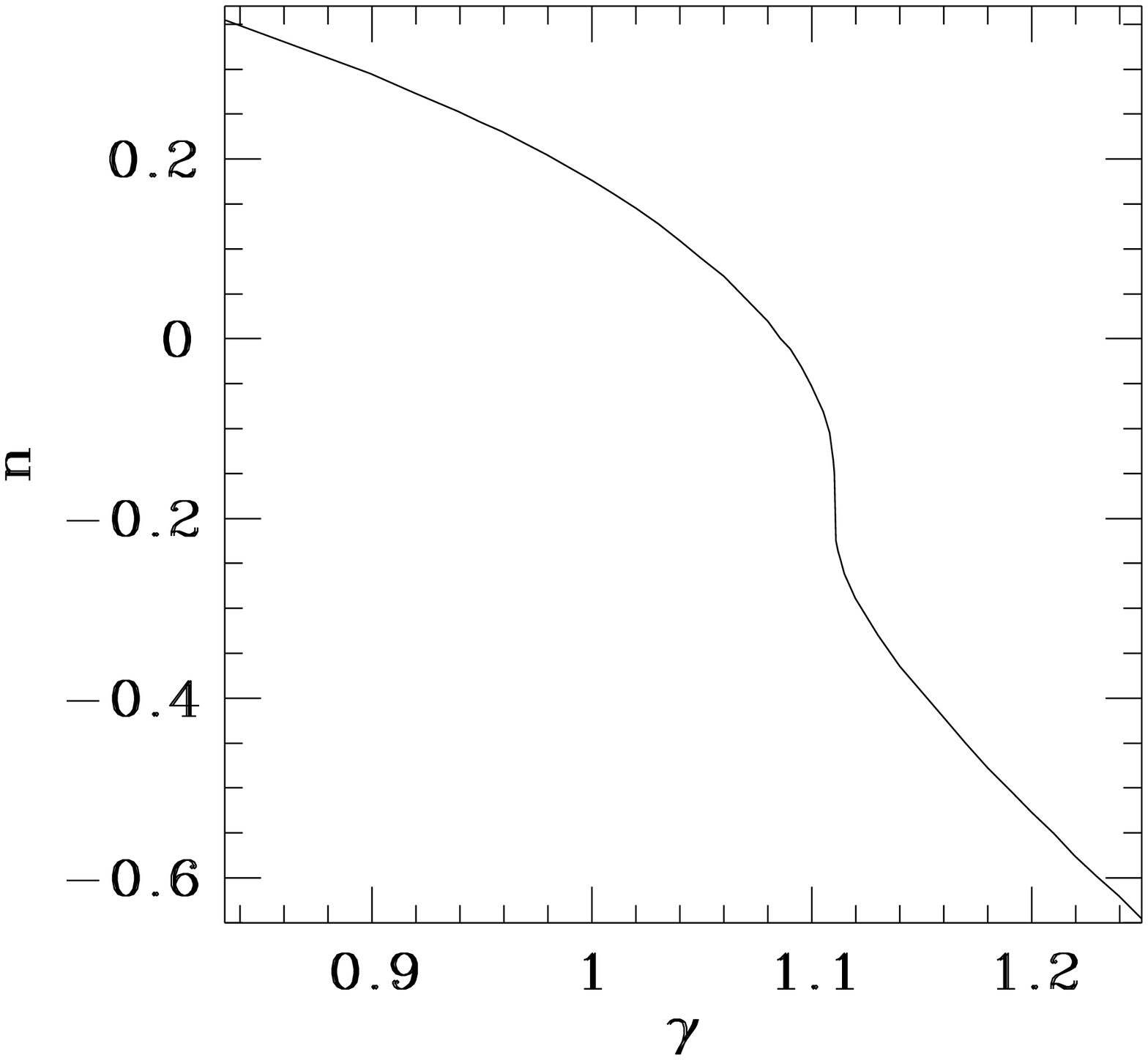}
\figcaption[linear]{Power of growth of the non-spherical deformation mode 
($m=2$). The amplitude of deformation ${\cal E}$ evolves as $\propto \rho^{n}$. 
The real part of exponent $n$ is plotted as a function of effective adiabatic 
index $\gamma$. 
\label{fig:linear}}

\plotone{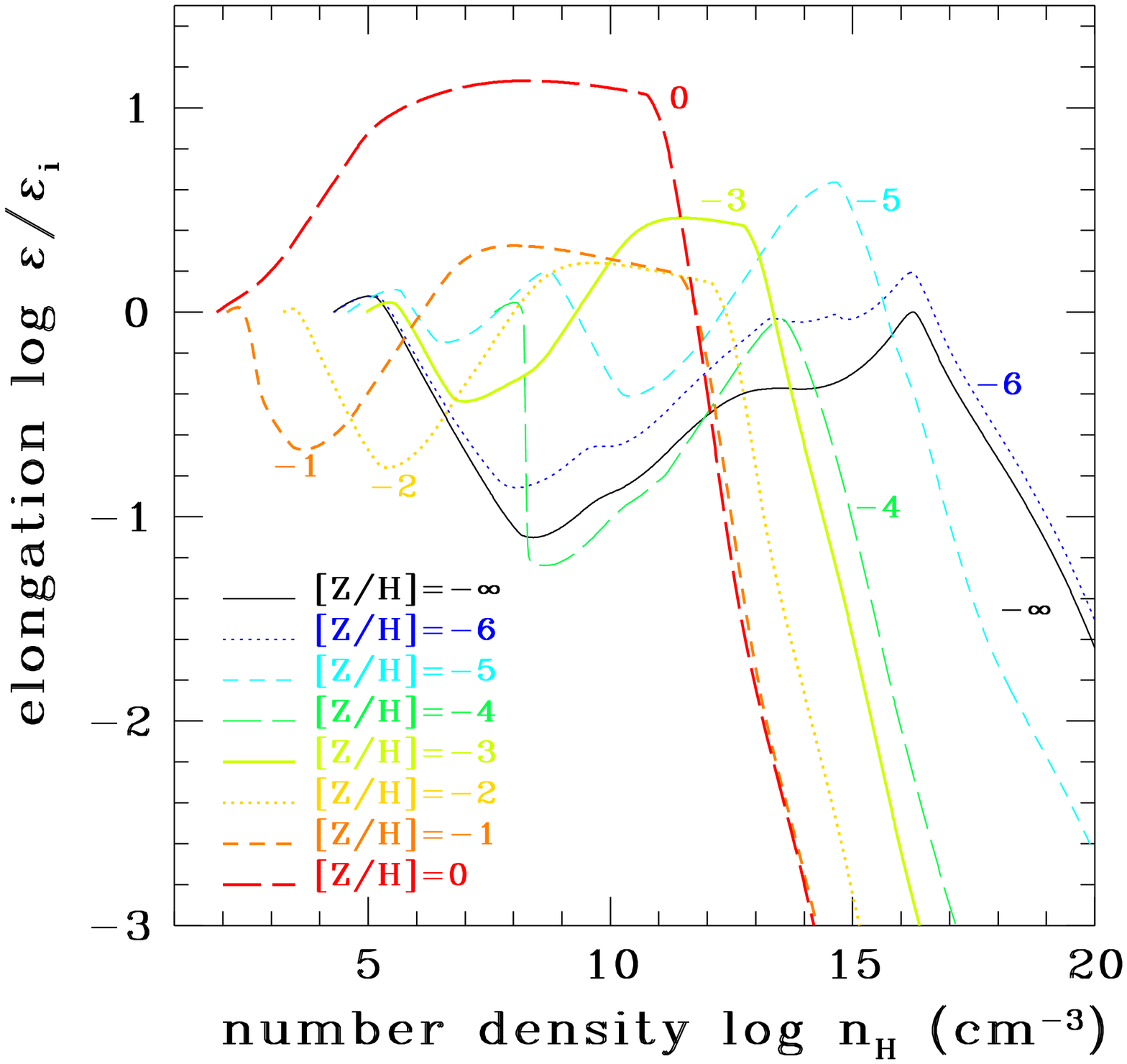}
\figcaption[elong]{Evolution of elongation ${\cal E}$ during 
the run-away collapse phase in the linear theory.
If the elongation exceeds a critical value, the cores are assumed to fragment
and the elongation is reduced as in the text.
This effect is not included in this Figure.
We assume that the run-away collapse begins after the first fragmentation, 
i.e., when the temperature begins to increase after the initial 
line-cooling phase.
\label{fig:elong}}

\plottwo{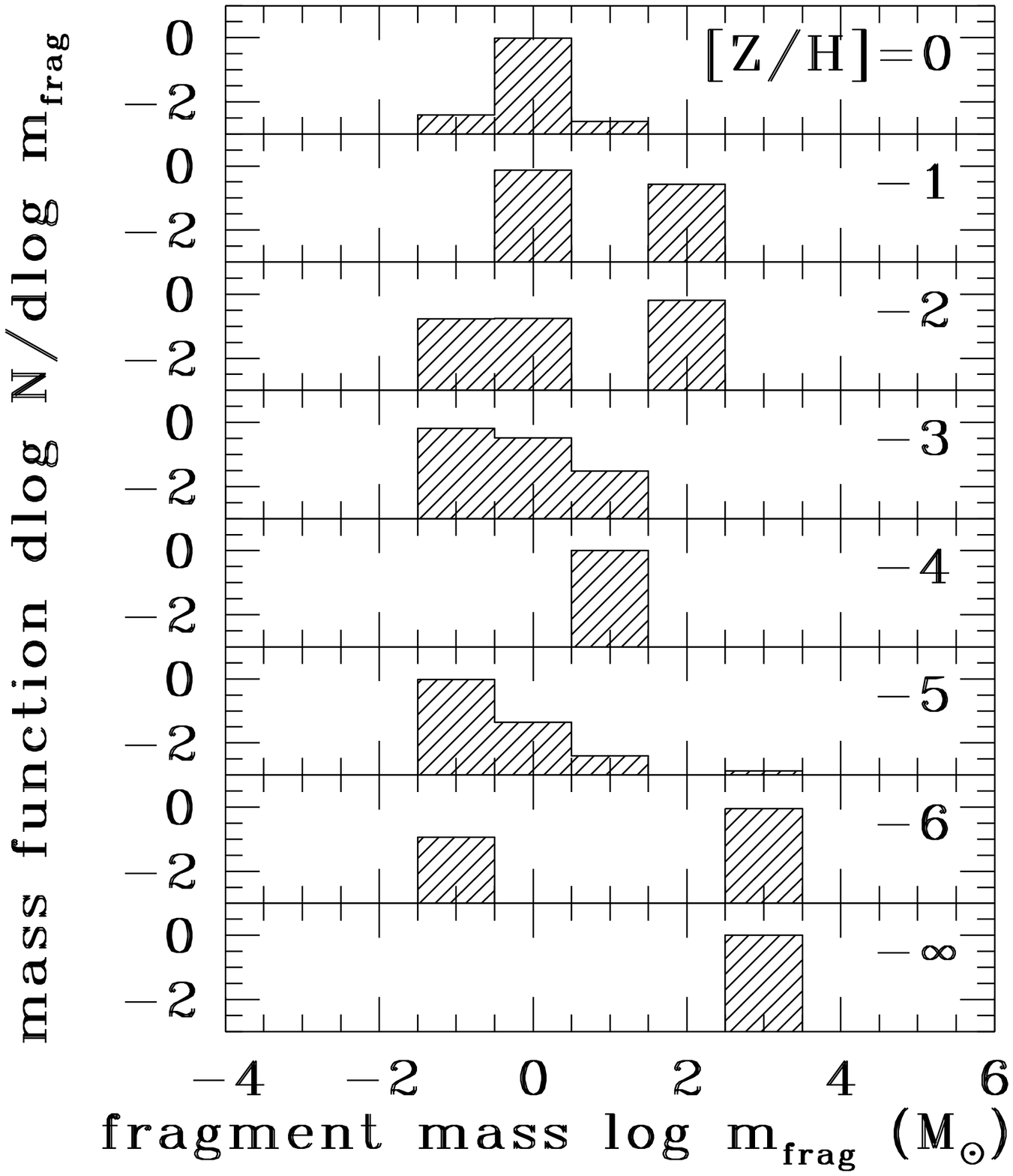}{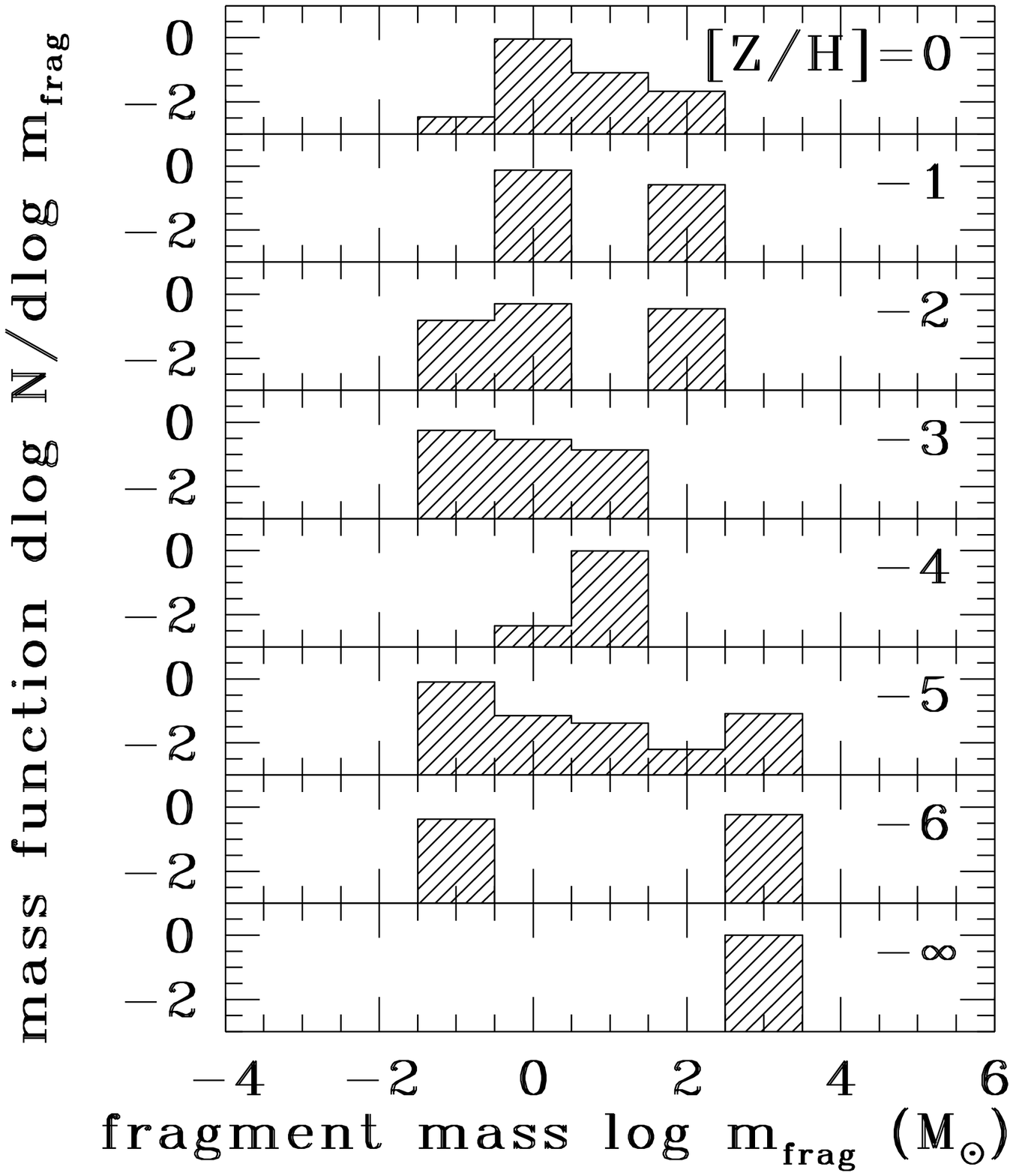}
\figcaption[fragmass]{The initial mass function of fragments for different
metallicities.
The number fraction in each bin is shown by the histogram.
Each bin has width $\Delta {\rm log}n_{\rm H}=1$ and
centered where values of ${\rm log}m_{\rm frag}$ are integer.
The initial distribution of ${\cal E}_{\rm i}/{\cal E}_{\rm NL}$ is 
(a)gaussian centered at 0.5 with variance $\sigma=0.1$ and 
(b) uniform between 0 and 1.
\label{fig:fragmass}}

\end{document}